\documentclass{aa}
\usepackage{graphicx}
\usepackage{txfonts}
\usepackage{natbib}
\usepackage{hyperref} 
\usepackage{dsfont}
\hypersetup{colorlinks=true,linkcolor=blue,citecolor=blue,urlcolor=blue}
\usepackage{orcidlink}
\usepackage{lscape}

\newcommand{\gaia}{{\it Gaia }}
\begin{document}






\title{2D  chemical evolution models II.} \subtitle{Effects of multiple spiral arm patterns  on O, Eu, Fe and Ba abundance gradients}

\author { E. Spitoni \orcidlink{0000-0001-9715-5727}\inst{1,2}  \thanks {email to: emanuele.spitoni@inaf.it} \and  G. Cescutti \orcidlink{0000-0002-3184-9918}   \inst{1,3,4}, A. Recio-Blanco \orcidlink{0000-0002-6550-7377} \inst{2}, I. Minchev  \orcidlink{0000-0002-5627-0355}\inst{5}, E. Poggio \orcidlink{0000-0003-3793-8505} \inst{2,6}, P. A. Palicio\orcidlink{0000-0002-7432-8709} \inst{2}, \\ F. Matteucci \orcidlink{0000-0001-7067-2302} \inst{1,3,4},  S. Peirani \inst{2},   M. Barbillon\inst{2}  \and A. Vasini \orcidlink{0009-0007-0961-0429} \inst{3}}
\institute{I.N.A.F. Osservatorio Astronomico di Trieste, via G.B. Tiepolo
 11, 34131, Trieste, Italy  \and Universit\'e C\^ote d'Azur, Observatoire de la C\^ote d'Azur, CNRS, Laboratoire Lagrange, Bd de l'Observatoire,  CS 34229, 06304 Nice cedex 4, France 
   \and Dipartimento di Fisica, Sezione di Astronomia,
  Universit\`a di Trieste, Via G.~B. Tiepolo 11, I-34143 Trieste,
  Italy
\and I.N.F.N. Sezione di Trieste, via Valerio 2, 34134 Trieste, Italy
\and Leibniz-Institut f\"ur
   Astrophysik Potsdam, An der
  Sternwarte 16, 14482, Potsdam, Germany 
   \and Osservatorio Astrofisico di Torino, Istituto Nazionale di Astrofisica (INAF), I-10025 Pino Torinese, Italy
 }

 \date{Received xxxx / Accepted xxxx}

\abstract {According to observations and numerical simulations, the Milky Way  could  exhibit several spiral arm modes with multiple pattern speeds, wherein the slower patterns are located at larger Galactocentric distances. }{Our aim is to quantify the effects of the spiral arms on the azimuthal variations of the chemical abundances for  oxygen, iron and for the first time  for neutron-capture elements (europium and barium)
 in the Galactic disc. We assume a model based on multiple spiral arm modes with different pattern speeds. 
The resulting model represents an updated version of previous 2D chemical evolution models. 
}{We apply new analytical prescriptions  for the spiral arms  in a 2D Galactic disc
chemical evolution model, exploring the possibility that
the spiral structure is formed by the overlap of chunks with different pattern speeds and spatial extent.  }{
The predicted  azimuthal
variations in abundance gradients  are dependent on the considered chemical element. Elements synthesised   on short time scales (i.e., oxygen and europium
in this study) exhibit larger abundance  fluctuations. In fact, for progenitors with short lifetimes, the chemical elements restored into the ISM  perfectly trace the  star formation perturbed by the passage of the spiral arms. The  map of the star formation rate  predicted by our chemical evolution model with multiple patterns  of spiral arms    presents arcs
and arms compatible with those revealed by multiple tracers  (young upper main sequence stars, Cepheids, and distribution of stars with low radial actions).
 Finally,  our model predictions are in  good agreement  with the azimuthal variations that emerged from the analysis of \gaia DR3 GSP-Spec [M/H] abundance ratios,  if at most recent times the pattern speeds match the Galactic rotational curve at all radii. 
}{We provide an updated version  of a 
2D chemical evolution model capable of tracing the azimuthal density
variations created by the presence of multiple spiral patterns  showing that elements synthesised on short time scales  exhibit larger abundance fluctuation.}

\keywords{Galaxy: disk -- Galaxy: abundances -- Galaxy: formation -- Galaxy: evolution -- Galaxy: kinematics and dynamics  -- ISM: abundances}

\titlerunning{Effects of multiple spiral arm patterns  on abundance gradients}

\authorrunning{Spitoni et al.}

\maketitle

\section{Introduction}

In various contemporary observational studies, significant azimuthal variations in the abundance gradients of external galaxies have been found. \citet{sanchez2015, sanchez_Me2016} extensively examined the chemical inhomogeneities of the external galaxy NGC 6754 using the Multi Unit Spectroscopic Explorer (MUSE) and concluded that the azimuthal variations in oxygen abundances are more prominent in the external regions of the galaxy.
 Using MUSE, \citet{vogt2017} conducted a study of the galaxy HCG 91c and found that the enrichment of the interstellar medium occurs primarily along spiral structures and less efficiently across them. \citet{li2013}  detected azimuthal variations in the oxygen abundance in the external galaxy M101.
\citet{ho2017} analysing the  galaxy NGC 1365, observed systematic azimuthal variations of approximately 0.2 dex over a wide range of radial distances that peak at the two spiral arms.

The investigation of azimuthal inhomogeneities of chemical abundances has also been carried out in the Milky Way system. \citet{balser2011,balser2015} and \citet{wenger2019}  studied the oxygen abundances of H II regions and found that the slopes of the gradients differed by a factor of two across their three Galactic azimuth angle bins. Additionally, significant local iron abundance inhomogeneities have been observed using Galactic Cepheids \citep{pedicelli2009,genovali2014}.
More recently, \citet{kovty2022}  analysed Cepheids from high-resolution spectra obtained by the Milky WAy Galaxy wIth SALT speCtroscopy project \citep[MAGIC,][]{kniazev2019},  to find that abundance asymmetries are particularly pronounced in the inner Galaxy and outer disc, where they reach approximately 0.2 dex, aligning with similar discoveries in nearby spiral galaxies.

 \citet{poggio2022} using  \gaia DR3 General Stellar Parametrizer - spectroscopy   (GSP-Spec, \citealt{recioDR32022b,recioDR32022a,vallenari2022})  showed statistically significant bumps on top of the observed radial metallicity gradients, with amplitudes  up to 0.05-0.1 dex. These results suggest that the assumption of a linear radial decrease is not applicable to this sample. The strong correlation between the spiral structure of the Galaxy and the observed chemical pattern in the younger sample suggests that the former could be responsible for the detected chemical inhomogeneities. The signature of the spiral arms is more prominent in younger stars and progressively disappears in cooler (and older) giants.

Several theoretical studies explored the nature and the origin of such azimuthal variations in the abundance gradients.  \citet{Khoperskov2018} focused on the formation of azimuthal metallicity variations in the disks of spiral galaxies, specifically in the absence of initial radial metallicity gradients. The findings indicate that the azimuthal variations in the average metallicity of stars across a spiral galaxy are not solely a result of the reshaping of an initial radial metallicity gradient through radial migration. Instead, they naturally emerge in stellar disks that initially possess only a negative vertical metallicity gradient. In \citet{khope2023}, they studied the influence of radial gas motions on the ISM metallicity near the spiral arms in the presence of an existing radial metallicity gradient. They found that the gas metallicity displays a dispersion of approximately 0.04 to 0.06 dex at a specific distance from the Galactic centre.

\citet[][hereafter ES19]{spitoni2D2018} presented one of the first 2D chemical evolution models capable to trace azimuthal variations. They showed that the main effect of considering      density fluctuations from the
 chemo-dynamical model by \citet{minchev2013} for the Galaxy  is  to create azimuthal variations of approximately 0.1 dex. Additionally, these variations are particularly noticeable in the outer regions of the Milky Way, in agreement with the recent findings in observations in external galaxies \citep{sanchez2015, sanchez_Me2016}.

Later, with their chemical evolution model in the presence of spiral arms, \citet{molla2019} predicted azimuthal oxygen abundance patterns for the last 2 Gyr of evolution are in reasonable agreement with recent observations obtained with VLT/MUSE for NGC 6754.

In ES19, it was shown that the amplitude of the azimuthal variation increases with the Galactocentric distance when the density fluctuation proposed by \citet{minchev2013} is considered;  as a consequence, different modes  with multiple spiral arm patterns coexist.
 If different modes combine linearly,  we  could approximate a realistic galactic disc by adding several spiral sets with different pattern speeds, as seen in observations \citep[e.g., ][]{meidt2009} and simulations \citep[e.g.][]{masset1997, quillen2011,minchev2012}.
 These patterns can include slow ones that are shifted towards the outer radii, as observed in studies such as \citet{minchev2006} and \citet{quillen2011}.  
It is important to point out that material spiral arms, propagating near the  co-rotation at all galactic radii, have been described by a number of recent numerical works with different interpretations (see \citealt{grand2012,comparetta2012,donghia2013, hunt2019}).

 To ensure a comprehensive perspective, it is important to emphasise that there is no agreement in the literature about the presence of various spiral arm modes exhibiting multiple pattern speeds.
Some authors claim that the spiral arms rotate like a rigid body with a single pattern speed \citep{lin1964,lin_shu1966}, while others suggest that the arms are stochastically produced by local gravitational amplification in a differentially rotating disk, with a process called "swing amplification" \citep{goldreich1965,julian1966}. It is also important to note that the morphology of the spiral structure in our Galaxy is highly debated, and no clear consensus has been reached, notwithstanding numerous efforts towards the mapping of its large-scale structure
\citep{georgelin1976,levine2006,hou2009,hou_han2014,reid2014, reid2019}.

In light of  the above considerations, in this article, we want to extend the work of ES19  focused on the effects of the spiral arm on the chemical enrichment of the Galactic thin disc,  by considering for the first time structures characterised by  multiple pattern speeds for different chemical elements, such as oxygen, iron, barium and europium.
Within this work, the terminology "thin and thick discs" refers to the low- and high-[$\alpha$/Fe] sequences in the [$\alpha$/Fe]-[Fe/H] plane. By defining the thin and thick discs based on morphology rather than chemical composition, a combination of stars from both the low- and high-[$\alpha$/Fe] sequences is identified, leading to a reciprocal identification as well \citep{minchev2015, Martig:2016ck}. Making this distinction is of utmost importance to prevent any confusion. Accordingly to ES19, we trace the chemical evolution of the thin disc component, specifically the low-$\alpha$ population. We assume that the oldest stars within this low-$\alpha$ component have ages of approximately 11 Gyr, which is consistent with asteroseismic age estimations \citep{victor2018,spitoni2019}.

Our paper is organised as follows: in Section \ref{ES19}, we summarise the chemical evolution model of ES19 in the presence of single pattern spiral arms. 
In Section \ref{sec_model_multi}, we present the methodology adopted in this paper to include in the chemical evolution model the density perturbations originated by spiral arms  with multiple patterns. In Section \ref{nucleo},  the adopted nucleosynthesis prescriptions are reported, and in Section  \ref{results_sec},
 we present our results and  in Section \ref{gaia_sec} we compare our results with \gaia DR3 observational data.
Finally, our conclusions and future perspectives  are drawn in Section \ref{conclu_sec}.

\section{The chemical evolution model of ES19 with single spiral patterns}
\label{ES19}
Here,  we provide some details of the 2D chemical evolution model presented by ES19. In particular,  in Section \ref{gas_accr} we recall the main assumption on the gas accretion history  and the adopted inside-out prescriptions of the Milky Way disc, whereas in Section \ref{spiral_es19} we present how the density fluctuations   created by  a single mode spiral arms  have been included in the chemical evolution model of ES19.

\subsection{The  gas accretion and inside-out prescriptions for the low-$\alpha$ disc} \label{gas_accr}

The Galactic thin disc  is assumed to be  formed by accretion of gas with pristine chemical composition \citep{matteucci1989} and the associated infall rate   for a generic
element $i,$ at the time $t,$ and Galactocentric distance $R$ (with no azimuthal dependence) is:
\begin{equation}\label{infall} 
\mathcal{I}(R,t,i)= \mathcal{X}_{i} \, a(R)  \,  e^{-\frac{t}{\tau(R)}},
\end{equation}
where $\mathcal{X}_{i}$ is the abundance by mass of the element
$i$ of the infall gas (that  is assumed
to be primordial here)       while 
the quantity $\tau_D(R)$ is the time-scale of gas accretion.
The coefficient $a(R)$ is constrained  by imposing a fit  
to the observed current total surface mass density $\Sigma_{D}$ profile.   We impose that the Galactic surface gas density of the disc at the beginning of the simulation (i.e. evolutionary time $t=0$ Gyr) is negligible.  
The observed present-day total disc surface mass density  in the solar neighbourhood is
$\Sigma_{\odot}=$ 54 M$_{\odot}$ pc$^{-2}$ (see \citealt{chiappini2001, romano2010, vincenzo2017, palicioAS2023}) and its variation with  the function of the Galactocentric 
distance reads:
\begin{equation}
\Sigma_D(R,t_G)=\Sigma_{\odot}e^{-(R-R_{\odot})/R_{D}},
\label{mass}
\end{equation}
where $t_G$ is the present time and
$R_{D}$   is the disc scale length which is assumed to be $3.5$ kpc. 
As suggested first by \citet{matteucci1989} and then by \citet{chiappini2001}, an important ingredient to reproduce the observed radial abundance gradients
along the Galactic disc  is the  inside-out formation on the disc: i.e.  the timescale $\tau_D(R)$  increases with the Galactic radius
assuming this linear relation:

\begin{equation}
 \tau_{D}(R) = 1.033 \cdot R \, [\mbox{kpc}] - 1.27 \, [\mbox{    Gyr}],
\label{tau}
\end{equation}
 The "inside-out" growth of the Galactic thin disc  has also been found in most zoom-in dynamical simulations in the cosmological context \citep{kobayashi2011,brook2012,bird2013,martig2014,vincenzo2020}.  We adopt the \citet{scalo1986} initial stellar mass function (IMF), assumed to be  constant in time and space. 
 
\subsection{Including  the effects of the density perturbations from a single spiral mode} \label{spiral_es19}
\citet{spitoni2D2018} presented a  new 2D chemical evolution model  designed to trace the azimuthal variations of  the abundance gradients  along the   disc, in particular showing the effects of spiral arm structures.
 The model divides the disc into  concentric shells 1 kpc-wide in the radial direction. Each annular region is composed by  36 zones of $10^{\circ}$ width each.  They showed the effects of spiral arms  on the chemical evolution considering  variations of the  star formation rate (SFR) along the different regions produced by density perturbations driven by  the analytical spiral arms  described by \citet{cox2002}. In particular,  they analysed the effects  of a  spiral arm structure characterised by a single mode, i.e. constant angular velocity pattern throughout the spiral structure.

Here, we  briefly summarise  the main model assumptions.
The expression for the change in the total mass density perturbation caused by spiral arms  given in an inertial reference frame that does not rotate with the Galactic disc  is 
\begin{equation}
\Sigma_S(R,\phi,t)= \chi(R,t_G) 
 M(\gamma).
 \label{sigma_s}
\end{equation}
The quantity $\chi(R,t_G)$ represents the present-day amplitude of the spiral density and  can be expressed as:
\begin{equation}
\chi(R,t_G)=\Sigma_{S,0} e^{-\frac{R-R_0}{R_{S}}},
\end{equation}
where $R_S$ is  the radial scale-length of the drop-off in
density amplitude of the arms,  $\Sigma_{0}$ is the surface arm
density at fiducial radius $R_0$.
In eq. (\ref{sigma_s}), the quantity $M(\gamma)$ is the modulation function  for the "concentrated arms" presented by   \citet{cox2002} and can be written as:

\begin{equation}
  M(\gamma)= \left(\frac {8}{3 \pi} \cos(\gamma)+\frac {1}{2} \cos(2\gamma) +\frac{8}{15 \pi} \cos(3\gamma)       \right)
  \label{MGAMMA}
\end{equation}
where $\gamma$ stands for

\begin{equation}
  \gamma(R,\phi,t)= m\left[\phi +\Omega_{s} t -\phi_p(R_0) -\frac{\ln(R/R_0)}{\tan(\alpha)} \right].
  \label{gamma_ref}
\end{equation}
In eq. (\ref{gamma_ref}), $m$  refers to the multiplicity (e.g.  the number of spiral arms),  $\alpha$ is
the pitch angle \footnote{In this model all the spiral arms have the same pitch angle $\alpha$. }, $\Omega_s$ is the  angular
velocity of the pattern,   $\phi_p(R_0)$ is the
coordinate  $\phi$  computed at $t$=0 Gyr and  $R_0$.
As underlined in ES19, an important feature of such a perturbation is that its
average density at a fixed Galactocentric distance $R$ and time $t$  is zero.

In the ES19 model, spiral arm overdensities are included in the chemical evolution  as perturbations of the \citet{kenni1998}  SFR law  (with the exponent $k$  fixed to 1.5) through   the following equation:
\begin{equation}
  \psi(R,t,\phi) = \nu \Sigma_g^k(R,t) \cdot  \delta_S(R,t,\phi)^k.
  \label{SFR2}
\end{equation}
 where $\nu$ is the star formation efficiency and  $\delta_S$ is an adimensional perturbation and defined as:

  \begin{equation}
    \delta_S(R,\phi,t) =1 + \frac{
      \chi(R,t_G)}{\Sigma_D(R,t_G)} M(\gamma),
  \label{delta_ref}
   \end{equation}
where $t_G$ is the present-day evolutionary time and having assumed  that the ratio $\chi(R,t)/\Sigma_D(R,t)$ is constant in time.
   More details and proprieties of the above-introduced expressions can be found in ES19.

\begin{figure}
 \includegraphics[scale=0.5]{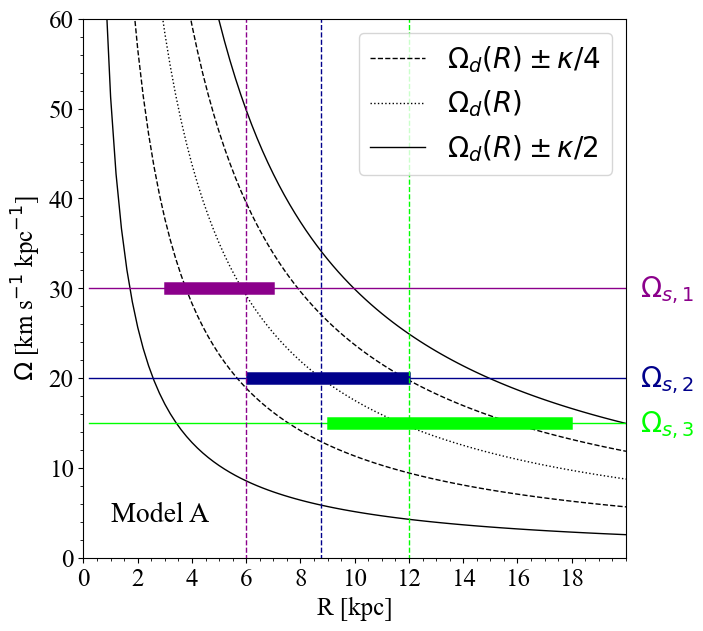}
  \caption{Spiral pattern speeds $\Omega_{s,1}(R)$, $\Omega_{s,2}(R)$ and $\Omega_{s,3}(R)$  of the  multiple spiral  modes moving at different pattern speeds (Model A in Table \ref{tab_models}) are indicated by the three coloured horizontal lines. 
  Inner and outer spiral structures (moving with the above-mentioned  pattern speeds) are also indicated  by the  thicker purple, blue and green lines, respectively.
    The disc angular velocity    $\Omega_d(R)$  computed by \citet{roca2014}  is indicated with the dotted line.
The 2:1 and 4:1  outer and inner Lindblad resonances (OLR and ILR) occur along the solid and dashed black curves, respectively. Resonances have been computed as $\Omega_{p2}(R) = \Omega_d(R) \pm  \kappa/2$ and $\Omega_{p4}(R) = \Omega_d(R) \pm  \kappa/4$, respectively where $ \kappa$ is the local radial epicyclic frequency. The long vertical dashed lines show the positions of the co-rotation radii assuming the three different $\Omega_S$ values.}
\label{patternA}
\end{figure}

\begin{table*}
\tiny \tiny
\caption{Properties of the three spiral arm structures with different pattern speeds  for the considered models. Pattern speed ($\Omega_{s,j}$)  limits of  Galactic spanned region ($R_{min}$ and  $R_{max}$ values)  and the co-rotation radius $R_{j, {\rm cor.}}$ for each modulation function $M_{MS,j}(\gamma_j)$ as defined in eq. (\ref{MSGAMMA}) are reported.
 All the listed Models are also characterised by:    the radial scale length of the drop-off in density 
amplitude of the arms $R_S$=7 kpc, disc scale-length $R_D$=3.5 kpc, surface arm density at $R_0$=8 kpc is $\Sigma_{S,0}$=20 M$_{\odot}$ pc$^{-2}$, star formation efficiency $\nu$=1.1 Gyr$^{-1}$, 
  and  multiplicity $m$=2.}
\label{tab_models}
\begin{tabular}{|c|cccc|cccc|cccc|}
\hline
  \hline
 & \multicolumn{4}{|c|}{$M_{MS,1}(\gamma_1)$}&  \multicolumn{4}{|c|}{$M_{MS,2}(\gamma_2)$}& \multicolumn{4}{|c|}{$M_{MS,3}(\gamma_3)$}\\
  & \multicolumn{4}{|c|}{}& \multicolumn{4}{|c|}{}& \multicolumn{4}{|c|}{}\\
Models & $\Omega_{s,1}$  &$R_{1,{\rm min}}$&  $R_{1,{\rm max}}$&  $R_{1,{\rm cor.}}$ & $\Omega_{s,2}$  &$R_{2,{\rm min}}$&  $R_{2,{\rm max}}$&$R_{2,{\rm cor.}}$   &$\Omega_{s,3}$  &$R_{3,{\rm min}}$&  $R_{3,{\rm max}}$&$R_{3,{\rm cor.}}$ \\
 & [km
s$^{-1}$ kpc$^{-1}$]&[kpc]&  [kpc]& [kpc]&[km
s$^{-1}$ kpc$^{-1}$]&[kpc]&  [kpc] & [kpc]&[km
s$^{-1}$ kpc$^{-1}$]&[kpc]&  [kpc]& [kpc] \\
 \hline 
 & \multicolumn{4}{|c|}{}&  \multicolumn{4}{|c|}{}& \multicolumn{4}{|c|}{}\\
A&30.0&3.0&7.0&6.0 &20.0&6.0&12.0&8.7&15.0&9.0&18.0&12.0\\ 
 & \multicolumn{4}{|c|}{}&  \multicolumn{4}{|c|}{}& \multicolumn{4}{|c|}{}\\
\hline
& \multicolumn{4}{|c|}{}&  \multicolumn{4}{|c|}{}& 
\multicolumn{4}{|c|}{}\\
A1&30.0&3.0&7.0&6.0&/&/&/&/&/&/&/&/\\
 & \multicolumn{4}{|c|}{}&  \multicolumn{4}{|c|}{}& \multicolumn{4}{|c|}{}\\
\hline
& \multicolumn{4}{|c|}{}&  \multicolumn{4}{|c|}{}& \multicolumn{4}{|c|}{}\\
A2&/&/&/&/&20.0&6.0&12.0&8.7&/&/&/&/\\
 & \multicolumn{4}{|c|}{}&  \multicolumn{4}{|c|}{}& \multicolumn{4}{|c|}{}\\
\hline
& \multicolumn{4}{|c|}{}&  \multicolumn{4}{|c|}{}& 
\multicolumn{4}{|c|}{}\\
A3&/&/&/&/&/&/&/&/&15.0&9.0&18.0&12.0\\
 & \multicolumn{4}{|c|}{}&  \multicolumn{4}{|c|}{}& \multicolumn{4}{|c|}{}\\
\hline
& \multicolumn{4}{|c|}{}&  \multicolumn{4}{|c|}{}& \multicolumn{4}{|c|}{}\\
B1&30.0&3.0&7.0&6.0 &20.0&6.0&12.0&8.7&17.0&9.0&18.0&10.2\\ 
 & \multicolumn{4}{|c|}{}&  \multicolumn{4}{|c|}{}& \multicolumn{4}{|c|}{}\\
\hline
& \multicolumn{4}{|c|}{}&  \multicolumn{4}{|c|}{}& 
\multicolumn{4}{|c|}{}\\
B2&30.0&3.0&7.0&6.0 &20.0&6.0&12.0&8.7&13.0&9.0&18.0&13.3\\ 
 & \multicolumn{4}{|c|}{}&  \multicolumn{4}{|c|}{}& \multicolumn{4}{|c|}{}\\
\hline
& \multicolumn{4}{|c|}{}&  \multicolumn{4}{|c|}{}& 
\multicolumn{4}{|c|}{}\\
A+C1& $\Omega_d$&3.0&7.0& [3-7]&$\Omega_d$&6.0&12.0& [6-12]&$\Omega_d$&9.0&18.0& [9-18]\\ 
(last 0.1 Gyr) &  & & &  & & & & & &&&\\

  & \multicolumn{4}{|c|}{}&  \multicolumn{4}{|c|}{}& \multicolumn{4}{|c|}{}\\
\hline
& \multicolumn{4}{|c|}{}&  \multicolumn{4}{|c|}{}& \multicolumn{4}{|c|}{}\\

A+C2& $\Omega_d$&3.0&7.0& [3-7]&$\Omega_d$&6.0&12.0& [6-12]&$\Omega_d$&9.0&18.0& [9-18]\\ 
(last 0.3 Gyr) &  & & &  & & & & & &&&\\
 & \multicolumn{4}{|c|}{}&  \multicolumn{4}{|c|}{}& \multicolumn{4}{|c|}{}\\
\hline 
& \multicolumn{4}{|c|}{}&  \multicolumn{4}{|c|}{}& \multicolumn{4}{|c|}{}\\
A+C3& $\Omega_d$&3.0&7.0& [3-7]&$\Omega_d$&6.0&12.0& [6-12]&$\Omega_d$&9.0&18.0& [9-18]\\ 
(last 1 Gyr) &  & & &  & & & & & &&&\\

 & \multicolumn{4}{|c|}{}&  \multicolumn{4}{|c|}{}& \multicolumn{4}{|c|}{}\\
\hline 
 \end{tabular}
\end{table*}
\section{Modeling multiple spiral patterns} \label{sec_model_multi}
Here, we extend the analysis of ES19 considering  the presence of multiple  spiral patterns and tracing their effects on the chemical evolution of diverse  elements synthesised at different time-scale, i.e. oxygen, iron, europium and barium. 
We consider  multi-pattern spiral arm structures  as suggested by \citet{minchev2016} to test on the chemical evolution models the  possibility that the spiral structure is composed by the overlapping of spatially limited clumps with different velocity patterns.
Analogously to eq. (\ref{sigma_s}), the expression for the time 
evolution  of the density perturbation, created by multiple pattern spiral arms is:

\begin{equation}
\Sigma_{MS}(R,\phi,t)= \chi(R,t_G) 
\sum_{j=1}^{N} M_{MS, j} \,(\gamma_j).
  \label{MS_equation}
\end{equation}
In the above expression,  $N$ is total number of spiral clumps and  the $M_{MS, j} \,(\gamma_j)$ term  is the new  modulation function  defined for the $j^{th}$ spiral mode clump  associated with the angular velocity $\Omega_{s,j}$ and can be   expressed  as follows:

\begin{equation}
  M_{MS, j} \,(\gamma_j) \equiv M_j(\gamma_j)\cdot \mathds{1}{\left[R_{j,\,\rm min},\  R_{j,\, \rm max}\right]},
  \label{MSGAMMA}
\end{equation}
where the value of the indicator function $\mathds{1}$ delimits the radial extension of the  considered spiral arm mode enclosed between the Galactocentric distances $R_{j,\,\rm min}$  and $R_{j,\,\rm max}$: is one if the argument is within the radial interval and zero otherwise.

 Imposing that    the ratio $\chi(R,t)/\Sigma_D(R,t)$ is constant in time, the  adimensional perturbation  $\delta_S$ defined in Section \ref{spiral_es19} becomes:

  \begin{equation}
    \delta_{MS}(R,\phi,t) =1 + \frac{
      \chi(R,t_G)}{\Sigma_D(R,t_G)} \sum_{j=1}^{N} M_{MS, j} \,(\gamma_j).
  \label{delta2}
   \end{equation}
As for the dimensional quantity introduced by ES19 (eq. \ref{delta_ref}) also  the new perturbation  defined in eq. (\ref{delta2}) has the  important feature that its average value at a fixed Galactocentric distance $R$ and time $t$ is   
  \begin{equation}
\langle\delta_{MS}(R,\phi,t) \rangle_{\phi} =1 + \frac{
      \chi(R,t_G)}{\Sigma_D(R,t_G)} \sum_{j=1}^{N}  \langle M_{MS, j} \,(\gamma_j) \rangle_{\phi}=1.
  \label{delta2_azim}
   \end{equation}

This prescription overcomes the too-simplified approach of ES19 taking into account the more  complex
behaviour already predicted  by N-body simulations 
\citep{quillen2011,minchev2012,sellwood2014}
 and external galaxies \citep{elme1992,rix1995,meidt2009}
where multiple spiral patterns have been found.

While ES19 explored this scenario by modelling individual spiral patterns, each with a different angular velocity, in this study we present a more self-consistent approach, considering simultaneously different pattern speeds and limited spatial extensions (as expected from observations and simulation) and using the same chemical evolution model for the Galactic disc. 


\begin{figure*}
\centering
 \includegraphics[scale=0.32]{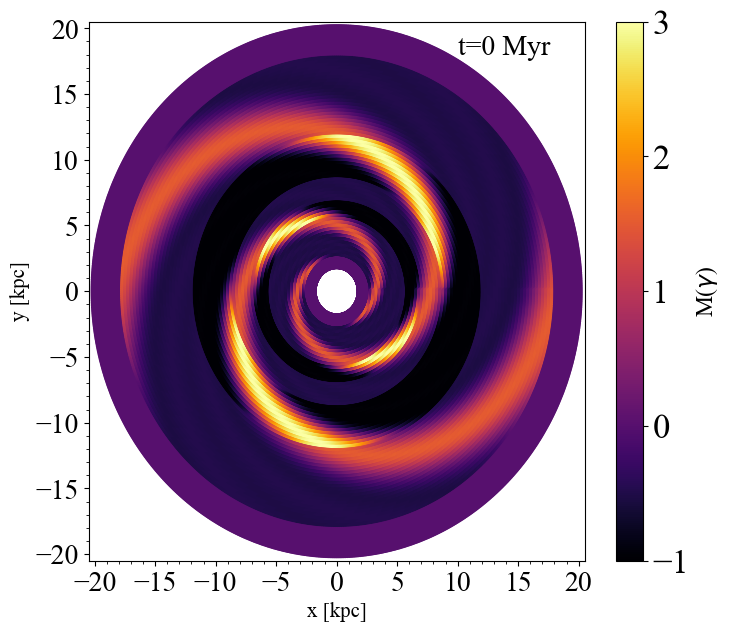}
 \includegraphics[scale=0.32]{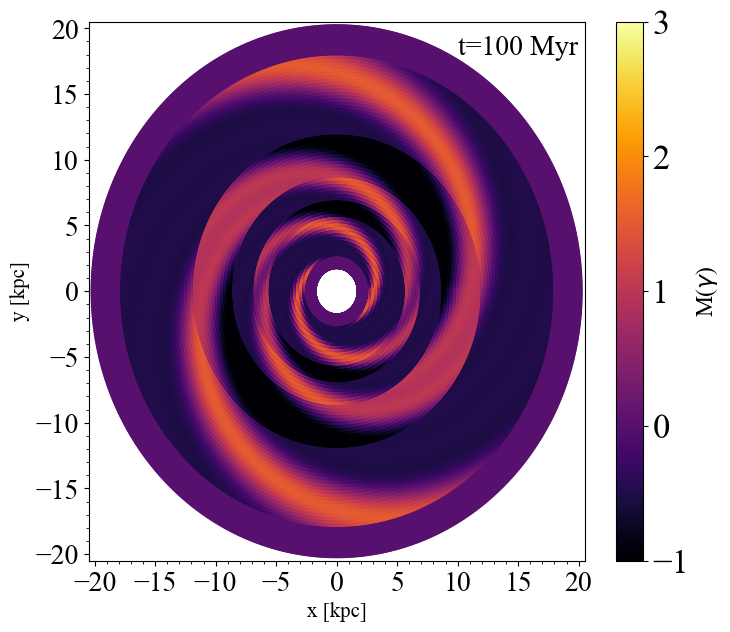}
 \includegraphics[scale=0.32]{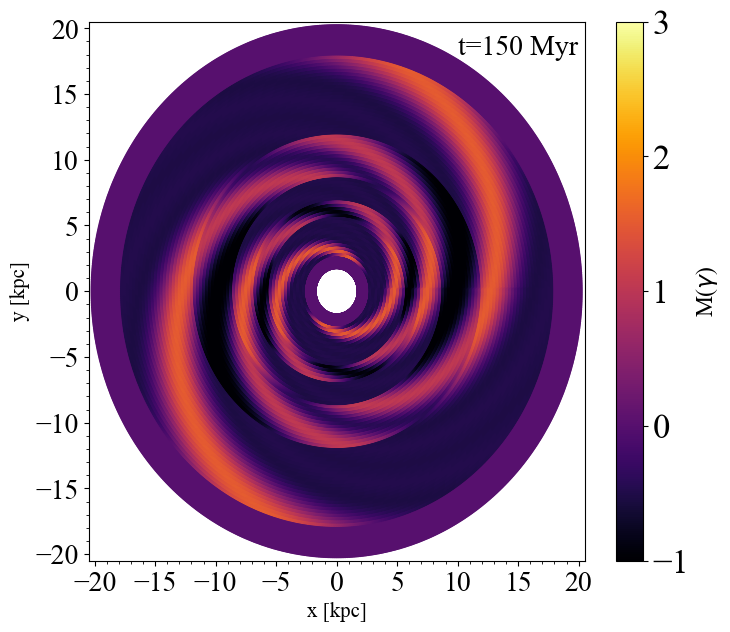}\\
 \includegraphics[scale=0.32]{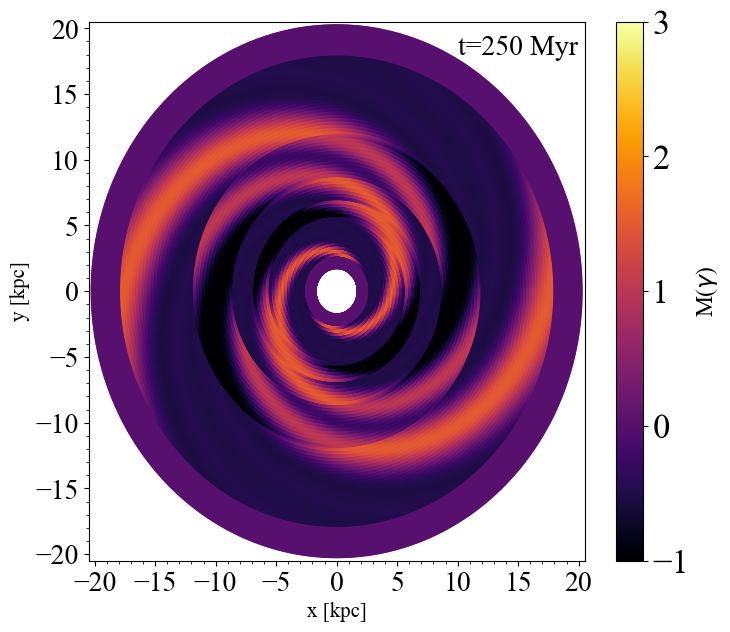}
 \includegraphics[scale=0.32]{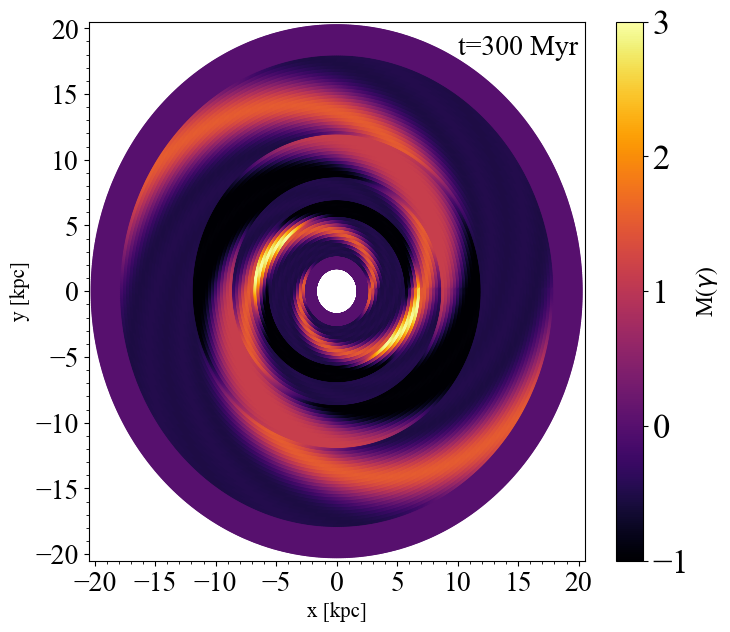}
 \includegraphics[scale=0.32]{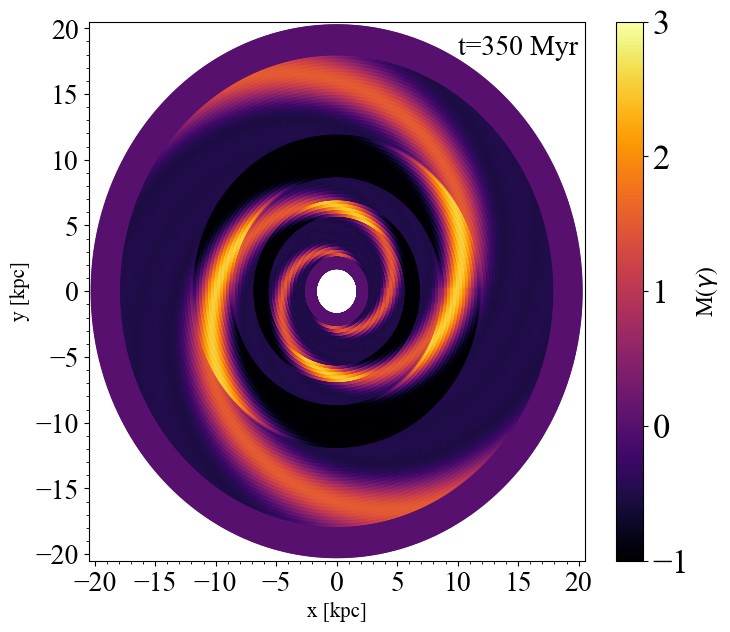}
  \caption{ Different snapshots of the temporal evolution  of the new modulation function $M(\gamma) \equiv \sum_{j=1}^{3} M_{MS, j}(\gamma_j)$ introduced in eq. (\ref{MSGAMMA}) for multiple spiral arm structures  assuming the same parameters as Model A  reported in Table \ref{tab_models} and also  characterised by multiplicity $m = 2$ spiral arms, fiducial radius $R_0 = 8$ kpc, pitch angle $\alpha=15 ^{\circ}$, and $\phi(R_0) = 0.$}
\label{modulation}
\end{figure*}

Following the same approach as \citet{minchev2016}, in Fig. \ref{patternA} we show the spiral pattern speeds $\Omega_{s,j}(R)$ of a spiral structure with multiplicity $m=2$ composed by  three  chunks moving at  different  pattern speeds $\Omega_{s,j}(R)$ (Model A in Table \ref{tab_models}).
  The $j^{th}$ spiral structures  is confined in the region   $ R \in \left[R_{j,\,\rm min},\  R_{j,\, \rm max}\right]$. 
  As in ES19, the disc rotational velocity $\Omega_d(R)$ has been extracted from the simulation
by  \citet{roca2014}.

The 2:1 and 4:1  outer and inner Lindblad resonances (OLR and ILR) have been computed as $\Omega_{p2}(R) = \Omega_d(R) \pm \kappa/2$ and $\Omega_{p4}(R) = \Omega_d(R) \pm \kappa/4$, 
respectively where $\kappa$ is the local radial epicyclic frequency.
 The velocity  of the central spiral  structure  is fixed at the value   $\Omega_{s,2}$ =20 km s$^{-1}$ kpc$^{-1}$ that is consistent with the  \citet{roca2014} model. A similar value was first estimated by
moving groups in the U-V plane by  \citet[][$\Omega$= 18.1$\pm$ 0.8 km s$^{-1}$ kpc$^{-1}$]{quillen2005}. 
It is interesting to note that one of the   co-rotational radii is located at  8.75 kpc, hence 
 close to the solar Galactocentric distance
 $(R,Z)_{\odot}=(8.249,0.0208)$ kpc \citep{gravity2021,bennett2019}. 
 We stress that  recent observations show  larger spiral  speed values in  the solar neighbourhood.  For instance, \citet{dias2019}  find a pattern speed of 28.2 km s$^{-1}$ kpc$^{-1}$ and \citet{quillen2018} of 29 km s$^{-1}$ kpc$^{-1}$ at 8 kpc. Nevertheless, to be consistent with the results presented in ES19,
we preferred  to retain both the  velocity disc and the co-rotation estimated in the vicinity of the solar system  ($\Omega_{s2}$)  by \citet{roca2014}.

As shown in ES19, the most significant effect of the spiral arms should take place at the co-rotation resonance where the chemical evolution should go much faster due to the lack of the relative gas-spiral motions and more efficient metal mixing. Moreover, it is widely established that discrete spiral waves in stellar disks can exist between their main resonances (ILR-OLR).
Since second-order resonances, i.e., 4:1 for a two-armed spiral  can also be quite important as shown in \citet{minchev2016} and giving rise to square orbits in the frame moving with the spiral pattern,  in Fig. \ref{patternA} they have been highlighted.

 We stress that also in \citet{castro-ginard2021} and \citet{quillen2018}, they find evidence for transient spiral arms, wherein different segments exhibit varying pattern speeds.  In this article, we will  refer to  spiral structures as suggested by \citet{minchev2016} and \citet[][see their Section \ref{hilmi_sec}]{hilmi2020} with velocity patterns rescaled to the rotational curve of 
 \citet{roca2014}.  
  Importantly, it should be noted that the methodology introduced in this work possesses versatility and can be expanded to analyse any generic velocity configurations within spiral arms and the disc.

In Fig. \ref{modulation}, we reported  different snapshots of the temporal  evolution     of the modulation function $M(\gamma) \equiv \sum_{j=1}^{3} M_{MS, j}(\gamma_j)$ introduced in eq. (\ref{MSGAMMA}) for Model A (see Table \ref{tab_models} for other parameter values). At the initial time, all the spiral chunks trace perfectly a  spiral arm structure with multiplicity $m=2$.  However,   as time goes by, different arc-shaped substructures become more and more prominent in the modulation function map  due to the different pattern speeds. It is possible to appreciate that as the number of arcs increases,    the amplitude of $M(\gamma)$ decreases.  
The strength of the perturbation is maximum when we recover the  $m=2$ spiral arm configuration after 350 Myr of evolution.

\section{Nucleosynthesis prescriptions}
\label{nucleo}
As anticipated in the Introduction, the main purpose of  this work is to  show the results of the azimuthal variations of
abundance gradients 
for oxygen, iron, europium and barium. In this Section, we provide the nucleosynthesis prescriptions for these elements.
\subsection{Oxygen and iron}
As done in a number of
chemical evolution models in the past 
\citep[e.g.][]{spitoni2019,spitoni2022,spitoni2023,vincenzo2019}, we adopt  for oxygen, iron,  the nucleosynthesis
prescriptions by \citet{francois2004} who selected the best sets of yields required to best fit the data
 (we refer the reader to their work for the details related to the observational data).
In particular, for Type II SNe yields, they found that the \citet{WW1995} values correspond to the best fit of the data.
This occurs because no modifications are required for iron yields,
as computed for solar chemical composition, whereas 
the best results for oxygen are given by yields computed as
functions of the metallicity. The theoretical
yields by \citet{iwamoto1999} are adopted for the Type SNeIa,  while the prescription
for single low-intermediate stellar mass is by 
\citet{van1997}.

 Although \citet{francois2004} prescriptions still
  provide reliable yields for several elements, we must be cautious
  about oxygen.
Several results have shown that rotation can affect the oxygen nucleosynthesis in massive stars
\citep{meynet2002} and, therefore, the chemical evolution  \citep{cescutti2010}, in particular at low metallicity. However, this does not affect our results since the data shown in this project are relatively metal-rich. Moreover, we are mostly interested in differential effects, rather than absolute values.
This set of yields has been widely used in the literature \citep{cescutti2007,cescutti2022,mott2013, spitoni2022,spitoni2023,palla2022} and turned out to be able to reproduce the main features of the solar neighbourhood. 

\subsection{Europium and barium}

Neutron star merger (NSM) is considered a fundamental production site for the Eu in our analysis. Following   \citet{matteucci2014} and \citet{cescutti2015},
the realization probability of double neutron star systems belonging to massive stars that will eventually merge, or simply the fraction of such events ($\alpha_{NSM}$).

They adopted a value of 2$\cdot$10$^{-6}$ M$_{\odot}$ for Eu yields. This is consistent with the range of yields suggested by \citet{koro2012}, who propose that NSM can produce from 10$^{-7}$ to 10$^{-5}$ M$_{\odot}$ of Eu per event. Moreover, it was assumed that a fixed fraction of massive stars in the 10-30 M$_{\odot}$ range are NSM progenitors. To match the present rate of NSM in the Galaxy (R$_{NSM}$=83$^{+209}_{-66}$ Myr$^{-1}$  \citealt{kalogera2004}),  the parameter $\alpha_{NSM}$ has been set to 0.05. The recent observation of the event GW170817 appears to support this rate \citep{matteucci2021,molero2021}.

We set a fixed time delay of 1 Myr for the coalescence of two neutron stars, consistent with the assumptions of \citet{matteucci2014} and \citet{cescutti2015}. We note that this model assumes all neutron star binaries have the same coalescence time, but a more realistic approach would consider a distribution function of such timescales, similar to the explosion time distribution for SNIa (see \citealt{simonetti2019} and \citealt{molero2021td}).
In this work, we do not consider the stochasticity  of the r-process events. Given the fixed and short delay time considered, the scenario is also compatible with other sources of r-process material such as MRD SNe \citep{winteler2012,nishimura2015}
 and collapsar \citep{siegel2019}.
We adopted the yields of \citet{cristallo2009,cristallo2011} for nucleosynthesis by s-process in low mass AGB stars (1.3 - 3 M${\odot}$), so in this work they play a role in particular for barium. The yields from non-rotating stars were utilised in our analysis, but they tend to overestimate the production of $s$-process elements at solar abundance. In contrast, the yields from rotating AGB stars produce insufficient neutron-capture elements. To address this issue, inspired by \citet{rizzuti2019} we divide the non-rotating yields by a factor of 2, to reproduce the observed data at solar metallicity.  $s$-process contribution from rotating massive stars has also been considered. Initially introduced by  \citet{cescutti2013,cescutti2014,cescutti2015} using the nucleosynthesis prescriptions proposed by \citet{frisck2012},  this study incorporates the yields by \citet{frisch2016}, as specified in Table 3 of \citet{rizzuti2019}.
\section{Results}
\label{results_sec}
In this Section,
we present 
the results of the effect of multiple spiral arm patterns  on  the chemical evolution of oxygen, iron, europium and barium in the  Galactic thin disc.
In Section \ref{results_A}, we  consider the spiral arm pattern speeds as shown in Fig. \ref{patternA} (Model A in Table \ref{tab_models}). In Section \ref{results_A1}, we will  report results for Models A1, A2 and A3 where the 3 spiral structures are considered separately (e.g. arms with single pattern speed) in different runs of the Galactic chemical evolution model (see Table \ref{tab_models} for further details). In Section  \ref{results_B} the effects of different angular velocities for  the most external spiral structure ($\Omega_{s,3}$) will be discussed (Models B1 and B2).

In Section \ref{hilmi_sec}, we  introduce additional complexities to the spiral
arm models presented so far,  considering spiral arms with different pattern speeds and modes.

Finally, in Section \ref{results_C}  we investigate the hypothesis that, in recent times, all Galactocentric distances are co-rotational radii, which means that the spiral arms are rotating at the same angular velocity as the Galactic disc lately.
In fact, several recent numerical studies have shown the possibility of  material spiral arms  propagating close to the co-rotation at various radii throughout the galaxy
\citep[see][]{grand2012, comparetta2012,hunt2019}.

All the model results to be presented in this paper adopt  the prescriptions from \citet{cox2002} spiral arms analytical model and also applied by ES19: the drop-off in density amplitude of the arms is fixed at a radial scale length of $R_S=7$ kpc,  the pitch angle is assumed to be constant in time and fixed at the value of $\alpha=15^\circ$. The surface arm density, $\Sigma_0$, is set to 20 M$_{\odot}$ pc$^{-2}$ (we refer the reader to ES19 for the motivation of this value) at the fiducial radius of $R_0=8$ kpc, and we also assume that $\phi_p(R_0)=0^\circ$.
It is worth mentioning that, as in ES19 we  follow the chemical evolution of the thin disc component, and we  assume that the oldest stars are associated with ages of $\sim$11 Gyr, which is in agreement with asteroseismic age estimates \citep{victor2018}.

\begin{figure}
 \includegraphics[scale=0.47]{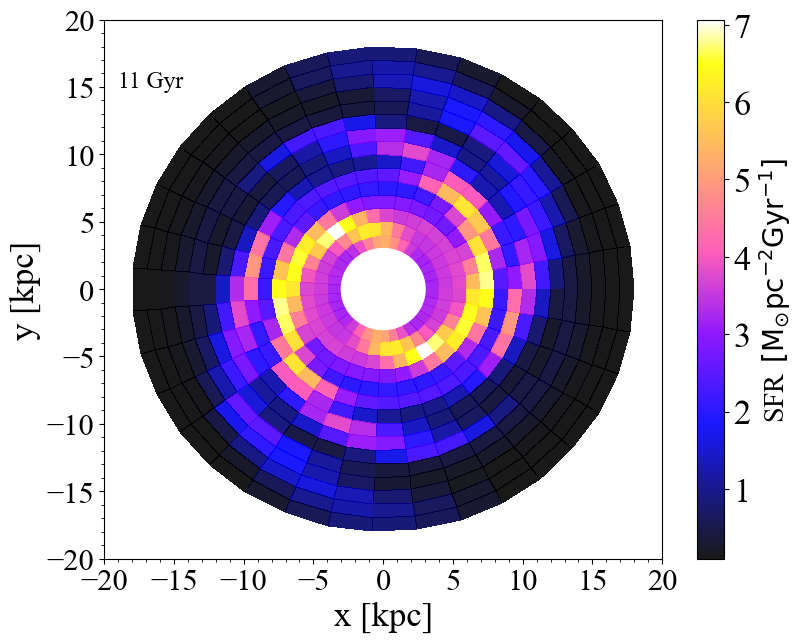}
   \caption{ 
   Galactic disc SFR resulting from Model A after 11 Gyr of evolution projected onto the Galactic plane after the start of disc formation (see Table \ref{tab_models} and text for model details). The colour-coding indicates the SFR values in units of M$_{\odot}$ pc$^{-2}$ Gyr$^{-1}$.  }
\label{SFR_11}
\end{figure} 
\begin{figure}
\centering
 \includegraphics[scale=0.25]{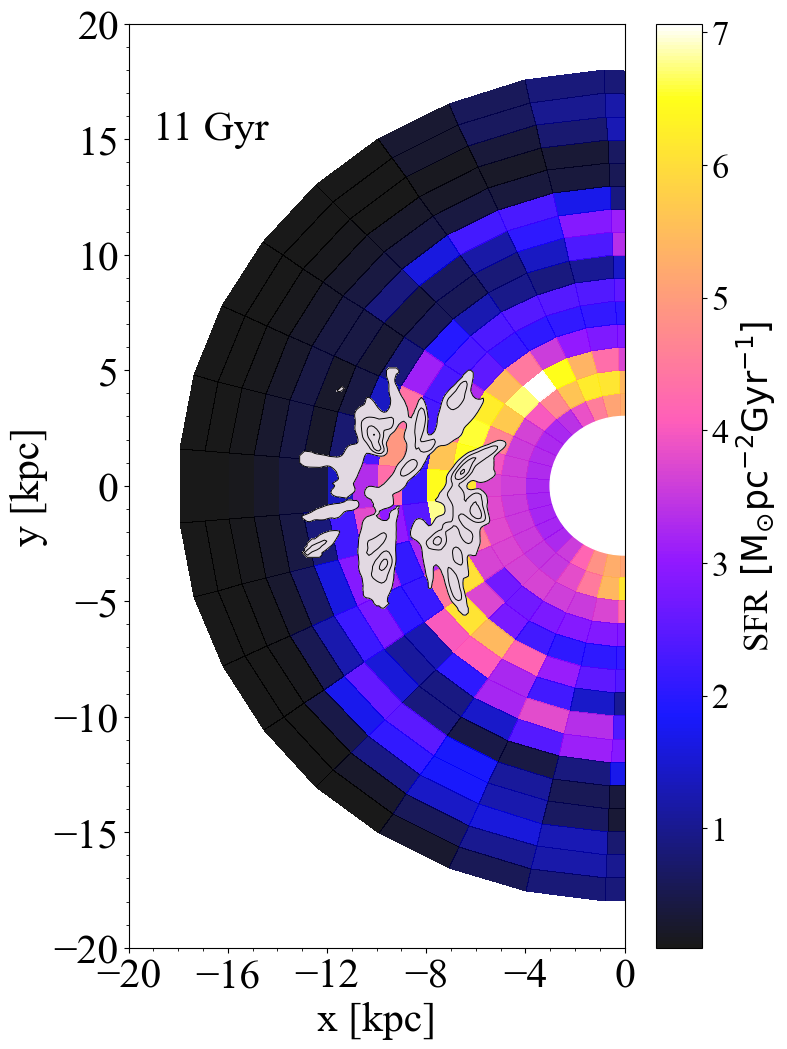}
 \includegraphics[scale=0.25]{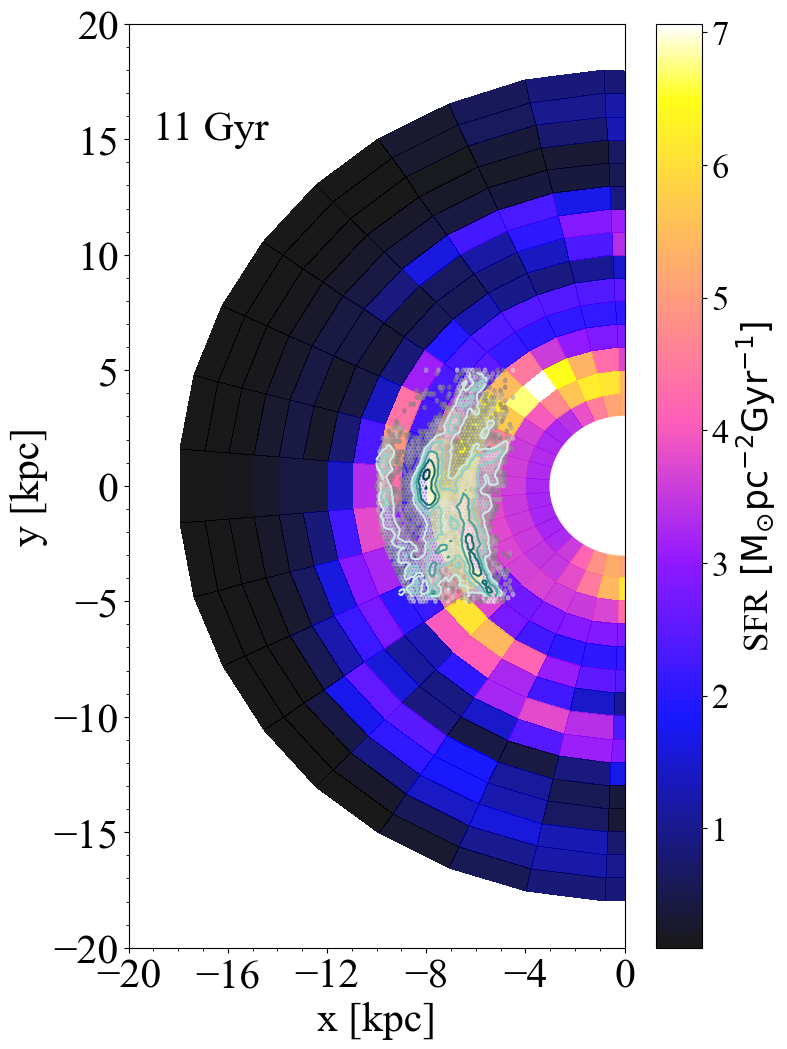}
  \includegraphics[scale=0.25]{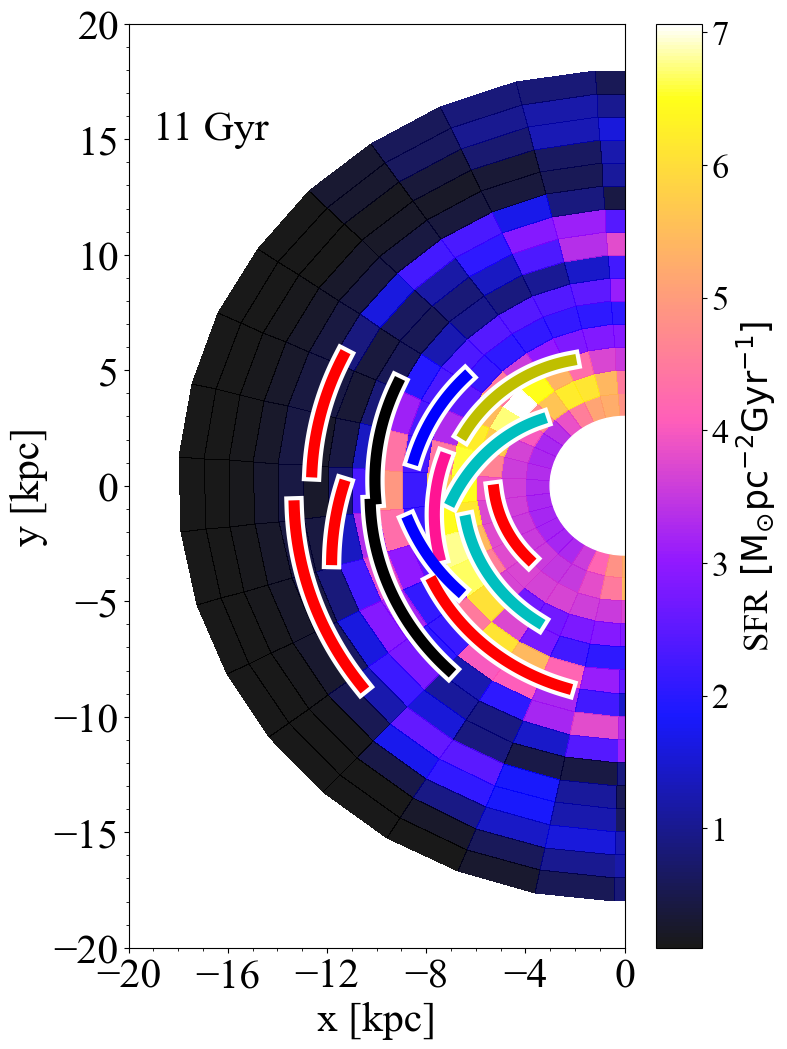}
   \caption{ A zoom-in  view of 2D map of the SFR resulting from Model A after 11 Gyr of the evolution  of Fig. \ref{SFR_11} compared  with some tracers of spiral arms or arcs  present
    in literature.   {\it Upper Panel:} 
   Overdensities in the distribution of UMS stars  of the \citet{poggio2018} sample    with  \gaia EDR3 astrometry  \citep{Egaia2021} as presented by  \citet{poggio2021}. We report  only positive values of  stellar overdensities as   defined by eq. (1) of \citet{poggio2021}.
    {\it Middle Panel:}   Distribution of the median  of  the radial action $J_R$   of Gaia DR3 stars (see text for more details) on the Galactic Plane (|$Z_{max}$| < 0.5 kpc and R<10 kpc) as computed by \citet{palicio2023} and here only reporting  the star with  $J_R<0.01$ $R_{\odot} V_{\odot}$ (green shaded area).
   {\it Lower Panel:}  Solid lines represent the segments of spiral arms  traced by Cepheids  of \citet{lemasle2022} in which the colour coding  is the one adopted  by \citet{palicio2023}.    
   }
\label{SFR_data}
\end{figure}

\begin{figure}
\centering
 \includegraphics[scale=0.45]{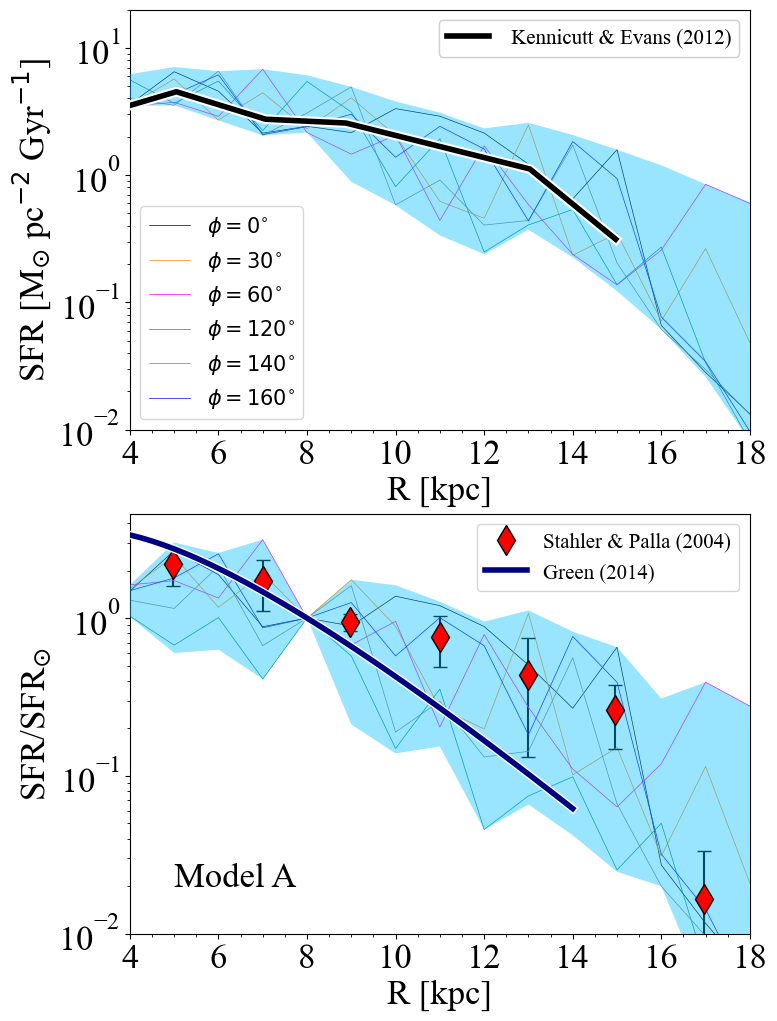}
   \caption{  Present-day radial SFR  profile  predicted by Model A for different azimuthals coordinates (colour lines).  In both panels, the shaded light-blue area denotes the range of maximum and minimum  SFR values at various Galactocentric distances.   {\it Upper Panel:} The predicted SFRs are compared with  the visible-band observations presented in \cite{guesten1982} and re-scaled by  \citet{kennicutt2012} to the total SFR of 1.9 M$_{\odot}$ yr$^{-1}$ of \citet{chomiuk2011}  as indicated by the black solid line. 
 {\it Lower Panel:}   Red diamonds with error bars are observational data for the star formation profile   from SN remnants, pulsars, and HII regions  normalised to the solar vicinity value (SFR/SFR$_{\odot}$) from \citet{stahler2004}.  The dark-blue solid represents the analytical fit of SN remnants compilation by \citet{green2014} as reported in \citet{palla2020} and \citet{spitoni2021}. The model predictions for different azimuths SFR$_{\phi}$ are  divided by their respective  solar vicinity SFR$_{\odot, \phi}$ values.
   }
\label{SFR_gradientA}
\end{figure}

\begin{figure*}
\centering
  \includegraphics[scale=.52]{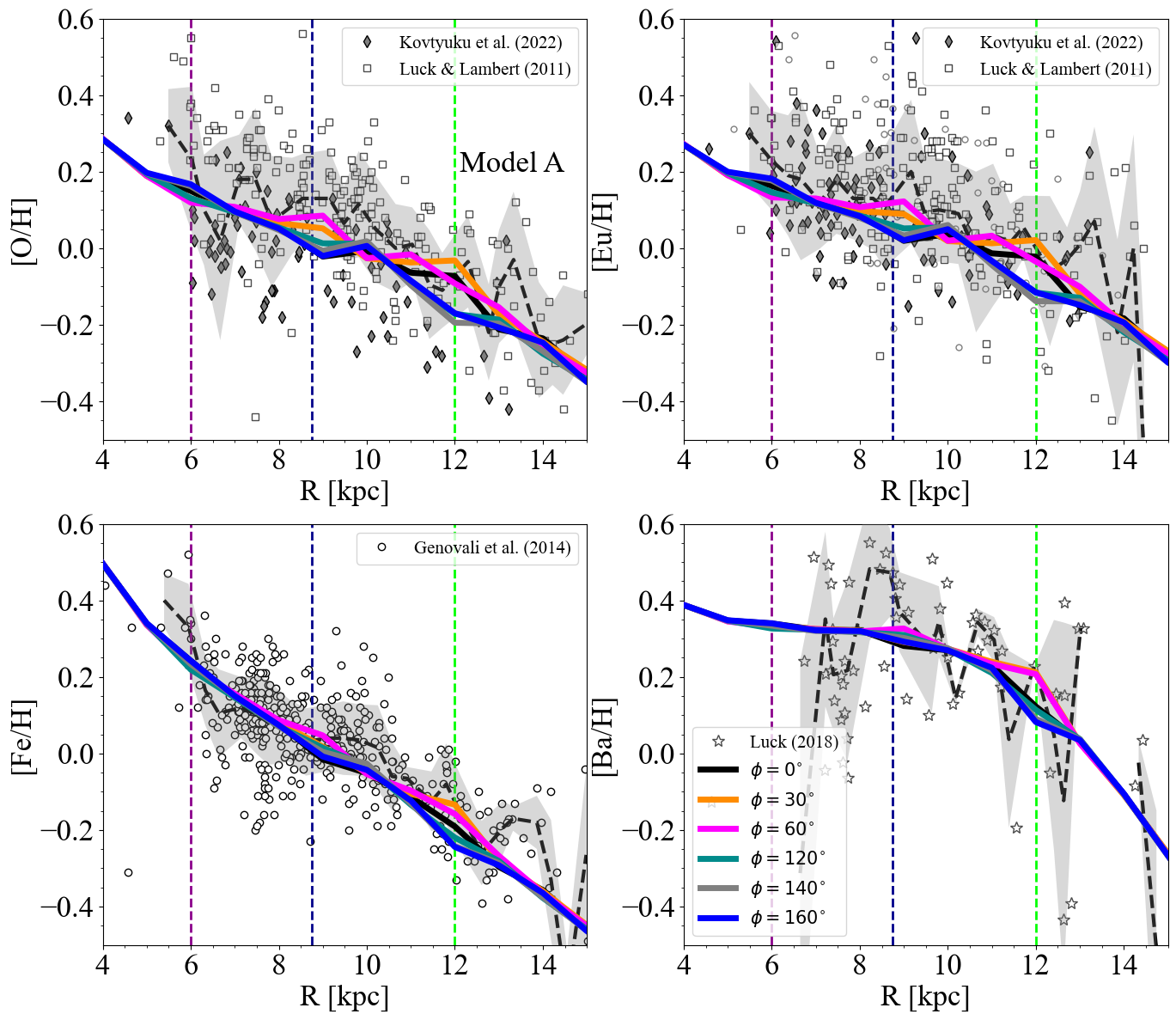}
  \caption{Present-day oxygen (upper left panel), europium (upper right panel), iron (lower left panel) and barium (lower right panel) abundance gradients for different azimuths as predicted by the Model A (see Table \ref{tab_models} for model details). In each panel, the coloured dashed vertical lines indicate the location of the co-rotation radii for the three  spiral structures characterised by different pattern speeds. Model predictions are compared with Cepheids data by \citet{kovty2022} and \citet{luck2011} for oxygen and europium, by \citet{genovali2014} for iron and by \citet{luck2018} for barium.    The data trends (black dashed lines) are computed by calculating the running median in bins of 0.5 kpc, with a 40\% overlap, and a minimum number of stars per bin of two for O, Fe and Eu, and one for Ba due to the low number of measurements in the sample. The shaded regions on the plot represent the standard deviation of these trends. For the sake of visualisation, we focus on the abundance range [-0.5, 0.6] dex, excluding stars with [Ba/H]>0.6 dex. } 
\label{gradient_A}
\end{figure*}

\subsection{Multi-pattern spiral arm structure: Model A}\label{results_A}
In Fig. \ref{SFR_11}, the predicted 2D map of the SFR projected onto the Galactic disc   by Model A after 11 Gyr of Galactic disc evolution (present-day) is shown. Although the considered spiral arms have  multiplicity  $m=2$,  the signatures of  multiple arcs and substructures originated by different velocity patterns  are well visible in the regions with enhanced star formation.  In  the upper panel  of Fig. \ref{SFR_data},  the position of the spiral arms in the Galaxy as mapped by the overdensities of upper main sequence (UMS) stars of  \citet{poggio2021} is overplotted on the predicted presented day SFR by Model A (the same as Fig. \ref{SFR_11}). In this plot, we report the stellar overdensities as defined by eq. (1) of \citet{poggio2021}  for positive density contrast values.

\begin{figure*}
\centering
  \includegraphics[scale=.25]{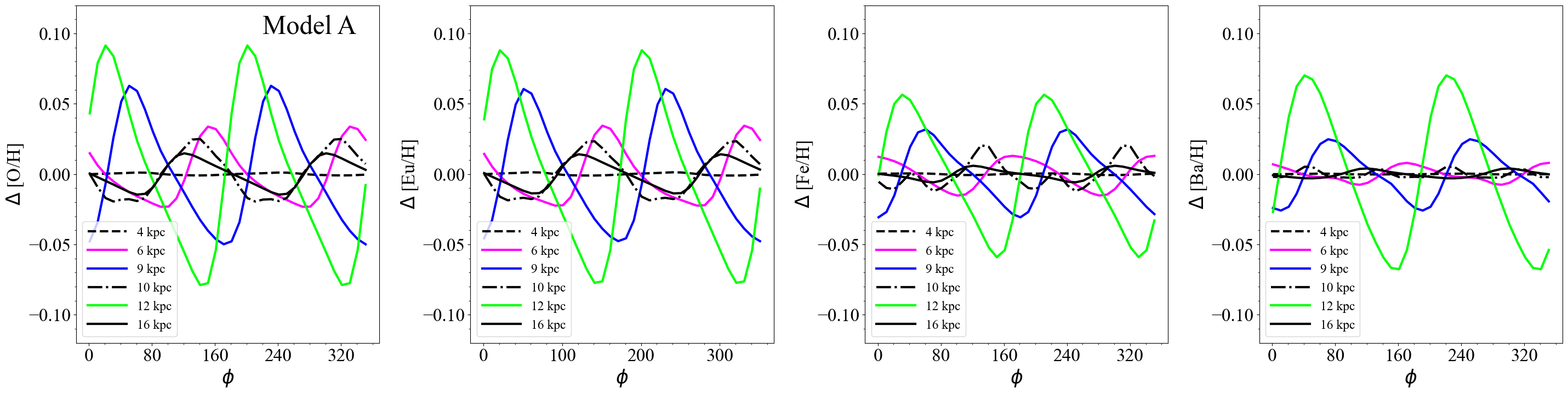}
  \caption{
Present-day residual azimuthal variations in oxygen, europium, iron and barium abundance, respectively  computed at different Galactocentric distances  for the multiple spiral structure  of Model A in Table \ref{tab_models}. The coloured solid  lines indicate the variations at the co-rotation radii for the three  spiral structures characterised by different pattern speeds with the same colour-coding as the dashed vertical lines  of Fig. \ref{gradient_A}.  
}
  \label{residual_A}
\end{figure*}
 
In the middle panel,  we show the map of the median  of  the radial action $J_R$ on the Galactic Plane (|$Z_{max}$| < 0.5 kpc and R<10 kpc) as computed by \citet{palicio2023}    for Gaia DR3 stars with full kinematic information only for bins with  $J_R<0.01$ $R_{\odot} V_{\odot}$,   which  trace the innermost region for the Scutum-Sagittarius spiral arms. For stars in the disc, $J_R$ can be interpreted as a parameter that quantifies the oscillation in the radial direction, with $J_R=0$ for circular orbits. We refer the reader to Appendix B of \citet{palicio2023} 
for a detailed explanation of how  the radial action $J_R$ has been computed.

Finally, in the lower panel, the  solid lines represent the segments of spiral arms  traced by Cepheids  in \citet{lemasle2022}. We note that there is   good agreement between the above-mentioned spiral arms tracers and the location of the predicted   enhanced star formation regions   driven by the passage of the multiple pattern spiral arm  of Model A.
In   Fig \ref{SFR_gradientA}, we note that the present-day  SFR  profiles predicted  by Model A  at different azimuthal coordinates throughout the Galactic disc are in agreement with the visible-band observations presented by
\ \citet{kennicutt2012}, \citet[][SN remnants, pulsars, and HII regions]{stahler2004}, and  \citet[][SN remnants]{green2014}.

In Fig. \ref{gradient_A}, we report for O, Eu, Fe and Ba  the present-day  abundance gradients predicted by Model A after 11 Gyr of evolution for different azimuthal coordinates.   It is important to stress that, as already found by \citet{cescutti2007}  and more recently by \citet{molero2023},  the predicted gradient  for barium is  almost  flat or slightly decreasing  in the innermost Galactic regions. 

As already underlined in the analysis of the azimuthal variation  of oxygen in ES19, the more significant  variations are located close to the  co-rotation. We note that as the   co-rotation is shifted towards the outer Galactic regions, variations  become more enhanced.
This  is in agreement with the results of ES19, where the density perturbation extracted by the chemo-dynamical model of \citet{minchev2013} was included.
In fact, also in this case 
significant variations have been found in chemical abundances in the outer Galactic regions. 

\citet{spitoni2D2018}  stressed that the chemical enrichment process at the  co-rotation  is expected to be more efficient because of the absence of relative gas-spiral motions. The  co-rotation radius experiences a higher star formation rate (SFR) due to increased gas overdensity, which persists for a longer duration. This leads to the formation of more massive stars and the ejection of more metals into the local interstellar medium (ISM).
To illustrate this, we represent in Fig.  \ref {residual_A} the excess of chemical abundances with respect to the azimuthal average.

\begin{figure}
\centering
  \includegraphics[scale=.4]{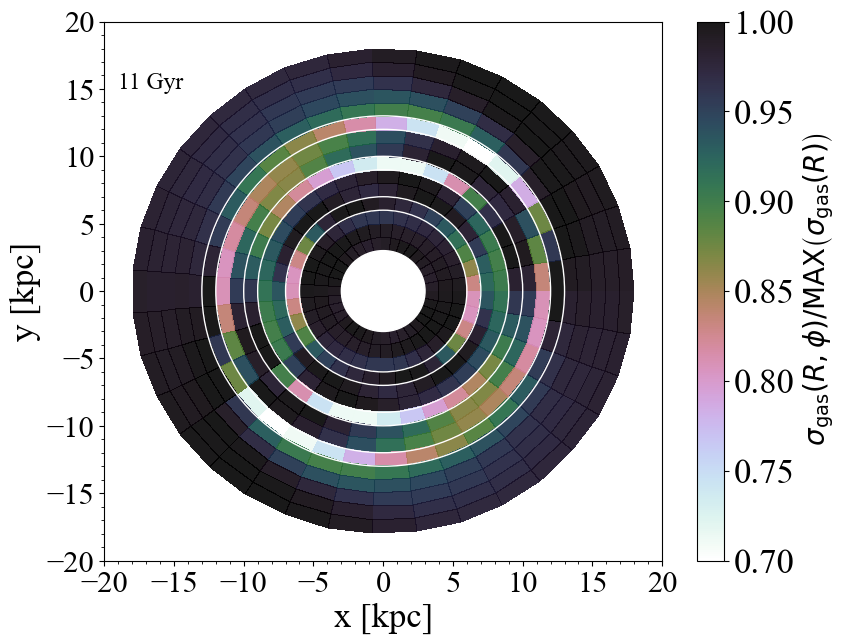}
\caption{
  Galactic disc  surface gas density distribution resulting from Model A after 11 Gyr of evolution normalised to the maximum values at different Galactocentric distances. We also highlight  with white edges  the  annular regions where three co-rotational radii are situated. 
 }
  \label{2D_gas_A}
\end{figure}

\begin{figure}
\centering
  \includegraphics[scale=.4]{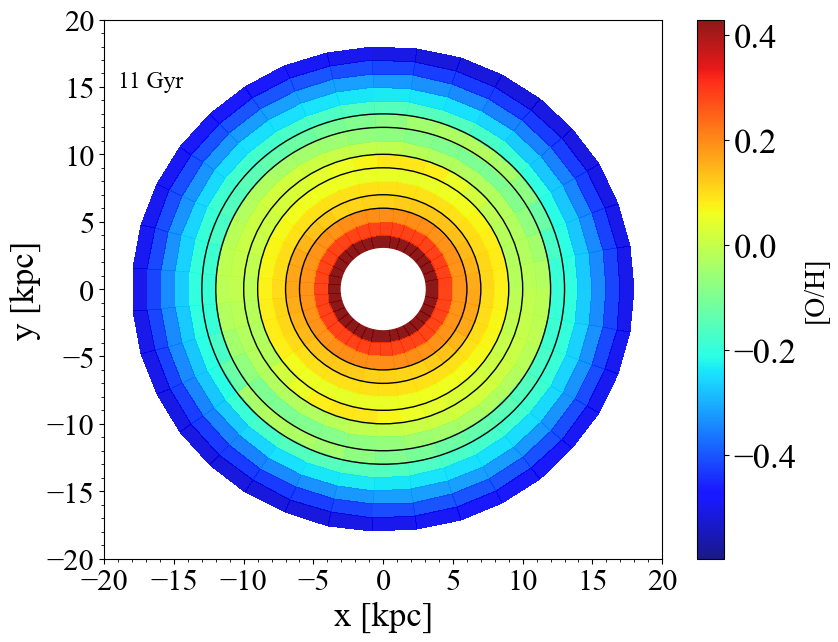}
  \includegraphics[scale=.4]{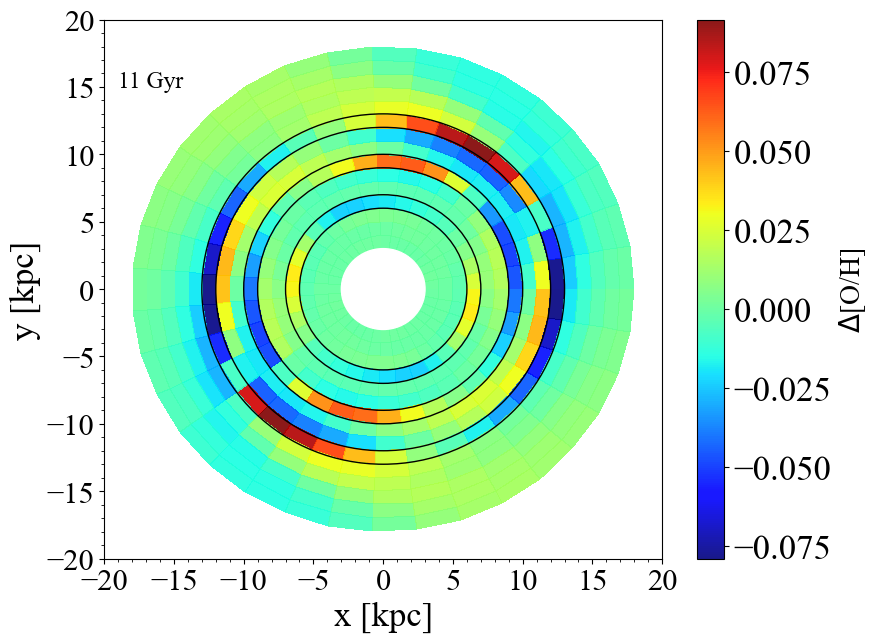}
  \caption{
  Galactic disc  oxygen abundances distribution resulting from Model A after 11 Gyr of evolution projected onto the Galactic plane after the start of disc formation (see Table \ref{tab_models} and text for model details).
 {\it Upper panel}: [O/H] values are reported. {\it Lower panel}: The distribution of residual azimuthal variations $\Delta$ [O/H] in oxygen are drawn. In each panel, we also highlight with black edges  the  annular regions where the three co-rotational radii are situated.  
}
  \label{2D_oxygen_A}
\end{figure}

With this model, we can  make predictions on the azimuthal variations originating from spiral arm structures for different chemical elements. From Fig. \ref{gradient_A}, it is clear that  azimuthal variations  depend on the studied chemical element: elements produced on short time-scales (i.e., oxygen  almost totally synthesised in Type II SNe  and europium  in   NSM  and via $r$-processes) show the largest variations. Since the progenitors in these cases have short lifetimes, the chemical elements restored into the ISM trace perfectly  the SFR fluctuations created by the spiral arm mass overdensities. It results  in a pronounced  variation in the abundance gradients compared with other elements ejected into the ISM after an important time delay. For example, the bulk of iron is produced by Type Ia SNe and  the timescale for  restoring it into the ISM depends on the assumed supernova progenitor model and the associate delay time distribution \citep{greggio2005,matteucci2009,palicioAS2023}. The typical timescale for the Fe enrichment in the Milky Way solar neighbourhood is around 1-1.5 Gyr \citep{matteucci2009}. 
  From Fig. \ref{gradient_A}, we note that a  larger spread in the chemical abundances  is  present  also in the Cepheid data for  elements produced in short time-scales (oxygen and europium)  compared to iron, in agreement with our model predictions. For the barium,  too few stellar abundances  are available  to make any firm consideration.

\begin{figure}
\centering
  \includegraphics[scale=0.45]{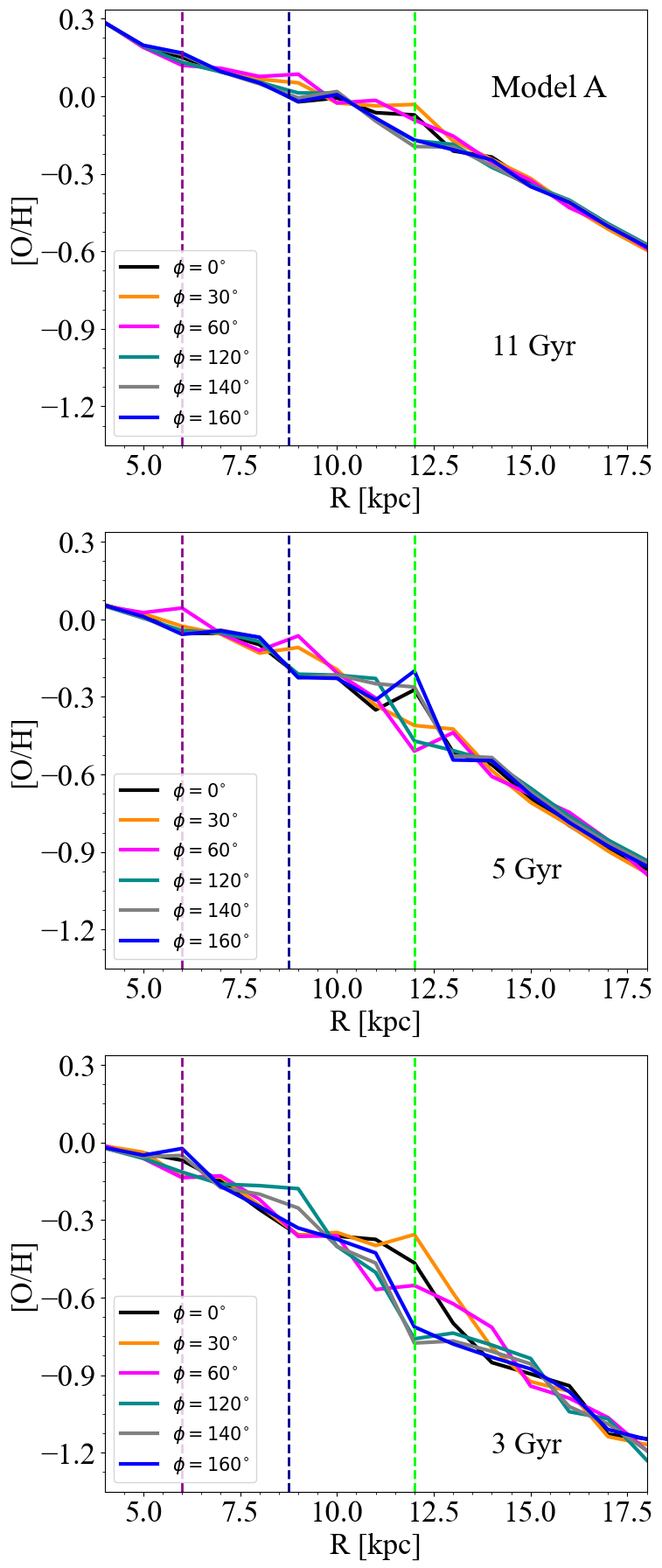}
   \caption{ Temporal evolution of the oxygen abundance gradient after 3, 5,  and 11 Gyr of thin disc evolution for different azimuth coordinates as predicted by Model A. Color code is the same as in Fig \ref{gradient_A}.
}
\label{agesA}
\end{figure}
In our approach we neglect  dynamical processes that could affect the gas distribution in the co-rotations.  For instance,  \citet{barros2021} pointed out,  from both hydrodynamic analytical solutions and simulations, that the interaction of the gaseous matter of the disk with the spiral perturbation should produce a flow of gas that establishes at the co-rotation region.  In particular, an inward flow of gas to the inner regions of the Galaxy and an outward flow to the outer regions are present  at the co-rotation circle, from which the flows diverge. As a natural consequence of this dynamic process, a ring-shaped void of gas should form at the co-rotation radius. 
Nonetheless,  \citet{lepine2017}  showed that  the Local Arm is an outcome of the spiral co-rotation resonance, which traps arm tracers and the Sun inside it. Hence,  it supports the scenario where some mass should cluster inside the co-rotation zones, thereby contributing to increased density in these  regions.
In conclusion,  two processes with opposite effects (gas depletion and clustering)  seem to coexist, and it is still   not clear which is the dominant one.

In our model, the lack of relative velocity between disc and spiral structures is the main cause of the pronounced azimuthal abundance variation at the co-rotation. As shown in Fig. \ref{2D_gas_A},  we have significant dips in gas distribution at  the three co-rotational radii.  However, these declines align exclusively with positive variations in chemical abundance (see Fig. \ref{2D_oxygen_A}).   In light of the  dynamical results mentioned by \citet{barros2021}, our findings must be considered  as an upper limit of the azimuthal variations originated by spiral arms.   On the other hand,  it is important to point out that  in the  chemodynamical simulations of the Milky Way-like spiral galaxies  as presented by \citet{khope2023}, there is no evidence  of any annular void region in the  gas distribution (see their Figure 2), which is a signature  of the presence of co-rotation, as suggested \citet{barros2021}. 

\citet{scarano2013}, analysing  external galaxies,  claimed  that    the presence of a step in metallicity  and  the change of slope of the gradient at this radius is due to the co-rotation. However, it is important to underline that a change in the slope in the abundance gradient can be the result of other chemo-dynamical processes such as  the inside-out formation scenario,   variable star formation efficiency or IMF throughout the Galactic disc  (see \citealt{matteucci2021} for a review).

In Fig. \ref{agesA}, we show the temporal evolution of the oxygen abundance gradients after 3, 5, 9 and 11 Gyr of evolution. At early times, the azimuthal variations are more prominent as already pointed out by ES19.

As the oxygen abundance increases (i.e. closer to the "saturation" level of the chemical enrichment),  the smaller  the chemical variations due to perturbations of the SFR are observed.
In addition, the Galactic chemical evolution is a cumulative  process in time. In early times, the stronger spiral structure induced azimuthal variations, which are later washed out by phase mixing.
Hence, we provide an important prediction for the high redshift galaxies with spiral arms   that will be  analysed in future works,  especially thanks to James Webb Space Telescope (JWST) discoveries.
In fact, \citet{fudamoto2022}, analysing the initial image captured by JWST of SMACS J0723.3-7327, highlighted  the presence of two   extremely red spiral galaxies  likely  in the cosmic noon  (1 < $z$ < 3).

\begin{table*}
\begin{center}
\tiny \tiny
\caption{ Maximum  azimuthal residual  variations $\Delta_{\rm max}$ [X/H] for oxygen, europium, iron and barium predicted by models B1 and B2 at the respective co-rotations of the outermost spiral chunk (characterised by  angular velocity  $\Omega_{s,3}$ and the co-rotation radius $R_{3,{\rm cor.}}$).   }
\label{tab_B12}
\begin{tabular}{|c|cc|cccc|}
\hline
  \hline
& \multicolumn{2}{|c|}{}&  \multicolumn{4}{|c|}{}\\
Models & $\Omega_{s,3}$&$R_{3,{\rm cor.}}$ & $\Delta{\rm max}$ [O/H]$_{3,{\rm cor.}}$ &$\Delta{\rm max}$ [Eu/H]$_{3,{\rm cor.}}$& $\Delta{\rm max}$ [Fe/H]$_{3,{\rm cor.}}$& $\Delta{\rm max}$ [Ba/H]$_{3,{\rm cor.}}$\\
 & [km s$^{-1}$ kpc$^{-1}$]&[kpc] &[dex] &[dex] &[dex] &[dex] \\
 \hline 
 & \multicolumn{2}{|c|}{}&  \multicolumn{4}{|c|}{}\\
B1&17.00&10.25&0.134&0.127 &0.075&0.069\\ 
 & \multicolumn{2}{|c|}{}&  \multicolumn{4}{|c|}{}\\
\hline
& \multicolumn{2}{|c|}{}&  \multicolumn{4}{|c|}{}\\
B2&13.00&13.30&0.186&0.177 &0.120&0.156\\ 
 & \multicolumn{2}{|c|}{}&  \multicolumn{4}{|c|}{}\\
\hline

 \end{tabular}
\end{center}
\end{table*}

\subsection{Single-pattern  spiral arms (Models A1, A2, and A3)}\label{results_A1}

Recent investigations pointed out that  it is very likely that the  Milky Way possesses  multiple modes with different patterns, with slower patterns situated towards outer radii \citep{minchev2006,quillen2011}.  
In ES19, only  the effects of  spiral arms  with single patterns in a chemical evolution model (e.g., considering  diverse velocities solely in different Galactic models) were presented. In order to be in agreement with the observations of external galaxies \citep{sanchez2015}, and with the results obtained using as fluctuations the ones form the chemodynamical model of \citet{minchev2013},  the authors assumed that the modes with different patterns combine linearly, and their total effects on
abundance azimuthal variations respond linearly to different modes considered.
 We confirmed this hypothesis  by testing that  the sum of  residual   azimuthal  variations predicted for Models A1, A2 and A3 (i.e. models where the 3 chunks of spiral arms are considered separately, see Table \ref{tab_models}) is almost identical to the ones of Model A:

\begin{equation}
\Delta\text{[O/H]}_{\text{A}} \simeq\Delta\text{[O/H]}_{\text{A1}}+\Delta\text{[O/H]}_{\text{A2}}+\Delta\text{[O/H]}_{\text{A3}}.
\end{equation}
We conclude that in our approach we do not find any amplifications of the  azimuthal variation in the  zones where the different chunks of spiral arms are connected and spatially overlap.

\subsection{Varying the pattern velocity for the external spiral structure: Models B1 and B2}
\label{results_B}
Models B1 and B2  (see Table \ref{tab_models}),  have the same pattern speed $\Omega_{s,1}$ and $\Omega_{s,2}$ and radial extension of the two innermost spiral structures as Model A, but  for the external clump  diverse velocities have been tested to  quantify the effects of different external co-rotational radii.   In Model B1,   the velocity of the outermost spiral chunk is  $\Omega_{s,3}=$ 17 km
s$^{-1}$ kpc$^{-1}$ (i.e. the  co-rotation is located at 10.25 kpc from the Galactic centre). On the other hand, in Model B2 we impose that $\Omega_{s,3}=$ 13 km
s$^{-1}$ kpc$^{-1}$ (i.e. a  co-rotation located at 13.30 kpc).

 In Table \ref{tab_B12}, we reported  
the maximum  residual azimuthal  variations $\Delta_{\rm max}$ [X/H] for oxygen, europium, iron and barium predicted by models B1 and B2 
at the respective co-rotations of the outermost spiral chunk  characterised by angular velocity $\Omega_{s,3}$.  We confirmed  that the lower  the velocity  $\Omega_{s,3}$ is, and the most prominent  amplitudes of the azimuthal variations are,  as already discussed in previous Sections and in ES19. 
It is evident  that a difference  in the pattern speed of $\Delta \Omega_{s,3}=4$  km
s$^{-1}$ kpc$^{-1}$  in the most external spiral mode leads to a shift of  the   co-rotation radius of roughly 3 kpc and has substantial effects on the chemical evolution of the Galactic disc.

\subsection{Spiral arms with different pattern speeds and modes  }
\label{hilmi_sec}

\begin{figure}
\centering
  \includegraphics[scale=.5]{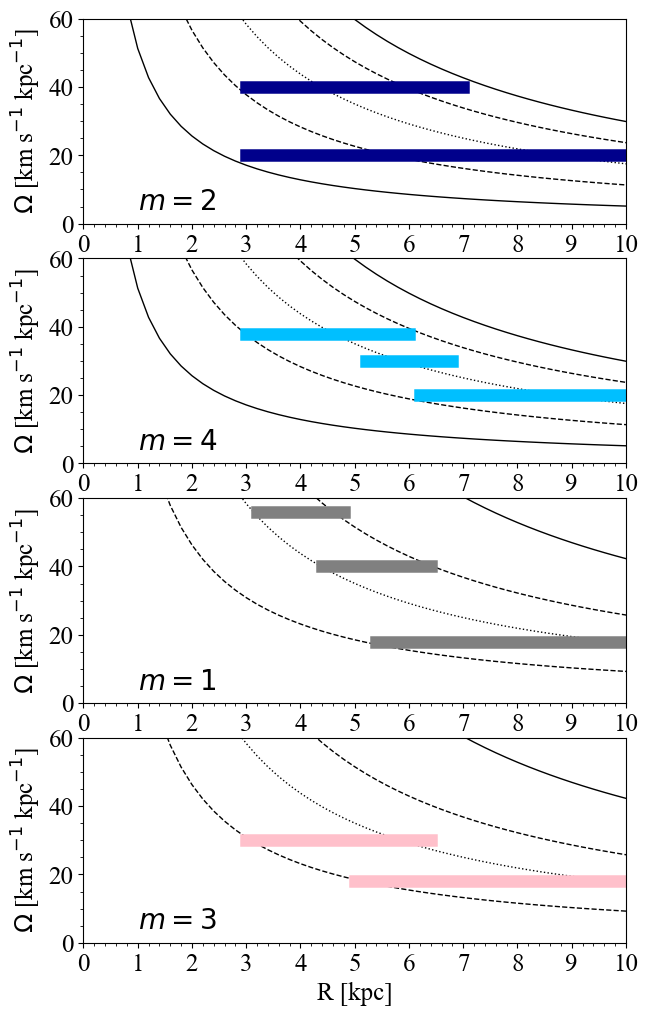}
   \caption{As Fig. \ref{patternA}, but for spiral modes extracted from the spectrogram analysis of Model1 by \citet{hilmi2020}, based on the \citet{buck_et_al2020} cosmological simulation. Different multiplicities are shown in each panel, as indicated. In each panel, the disc angular velocity $\Omega_d(R)$  is indicated with the dotted line. 
    In the  two  upper panels (with modes $m=2$ and $m=4$), we also show the resonances  $\Omega_{p2}(R) = \Omega_d(R) \pm  \kappa/2$ and $\Omega_{p4}(R) = \Omega_d(R) \pm  \kappa/4$,
   indicated with solid and dashed black curves, respectively. In the last two panels solid and dashed black  
have been computed as $\Omega_{p1}(R) = \Omega_d(R) \pm  \kappa$ and $\Omega_{p3}(R) = \Omega_d(R) \pm  \kappa/3$.}
  \label{roca_hilmi}
\end{figure}

In this Section, we introduce additional  complexities to the spiral arm models presented so far, with the aim of examining the influence of multiple patterns and different modes on the chemical evolution of the thin disc. We extracted the pattern speeds from  the power spectrogram constructed by \citet{hilmi2020}  of  the $m$ = 1, 2, 3, and 4 Fourier components using a time window of 350 Myr
for their Model 1. This model is based  on the high-resolution hydrodynamical simulations of MW-sized galaxies from the NIHAO-UHD project  of  \citet{buck_et_al2020} (galaxy g2.79e12).

\begin{figure}
\centering
  \includegraphics[scale=.3]{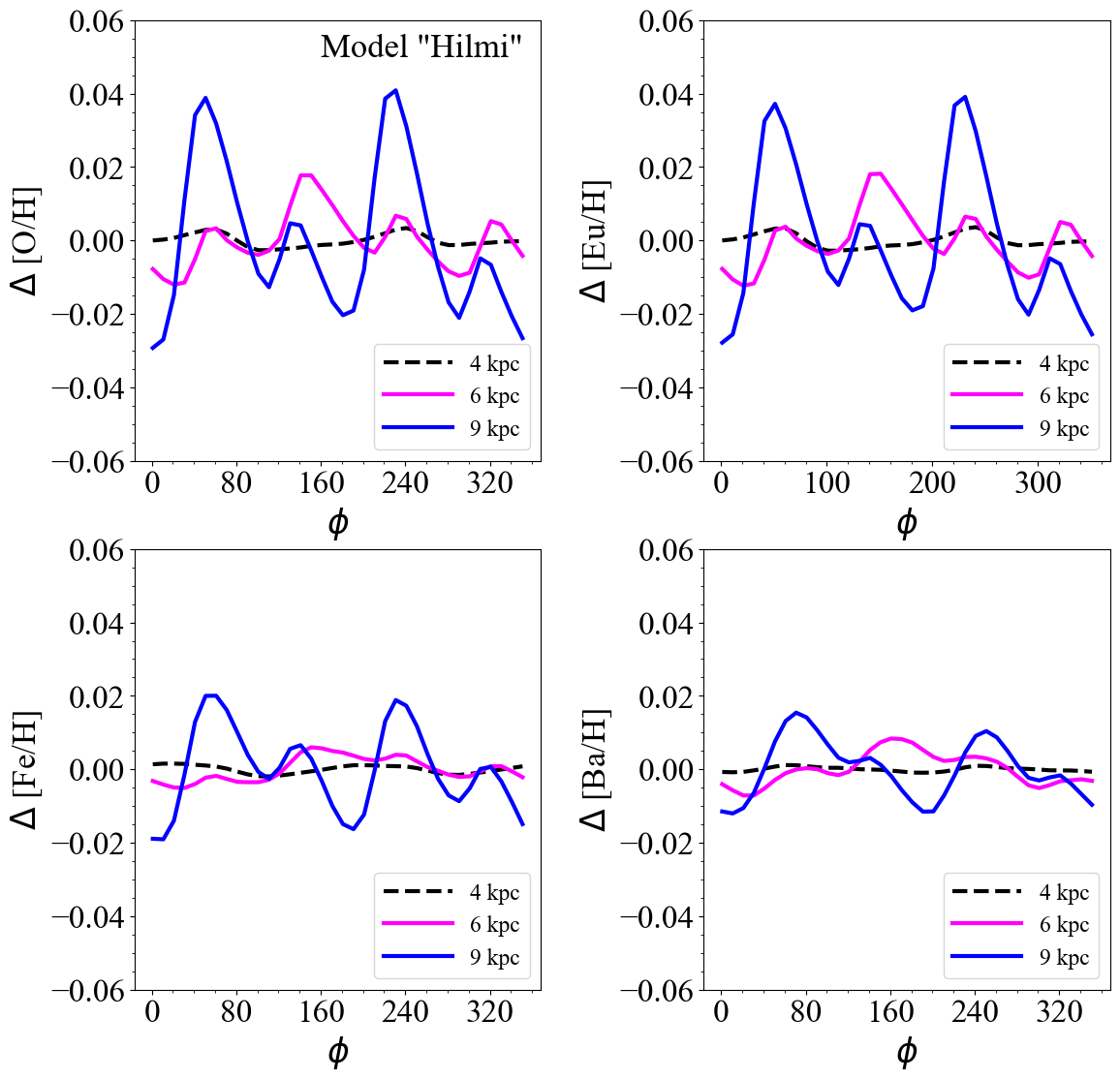}
   \caption{Present-day residual azimuthal variations in oxygen, europium, iron and barium abundance, respectively computed at 4, 6 and 9 kpc for  Model1 by \citet{hilmi2020}, as introduced in Section \ref{hilmi_sec}, where the spiral structure is  characterised multiple modes of different multiplicity from 1 to 4. } 
  \label{d_hilmi}
\end{figure}

In  Fig. \ref{roca_hilmi}, we present the velocity patterns for various modes extracted by \citet{hilmi2020} and scaled to the circular velocity determined by \citet{roca2014}.  
 In analogy with  eq. (\ref{MS_equation}),  the expression for the time-evolution of the density perturbation created by different speeds and modes of Fig. \ref{roca_hilmi} can be written as:
\begin{equation}
\Sigma_{MS}(R,\phi,t)= \chi(R,t_G) \sum_{m=1}^{4} \left( A_{m}
\sum_{j=1}^{N_m} M_{MS_{m}, j} \,(\gamma_j)
\right),
\end{equation}
 where $N_m$ is the number of spiral clumps associated with the mode $m$. The coefficients  $A_m$   are set by adopting the power spectrograms from figure 11 of \citet{hilmi2020}   to redistribute the spiral density perturbation across different modes ($A_1$=0.1, $A_2$=0.4, $A_3$=0.1 and $A_4$=0.4). 
In Fig. \ref{d_hilmi}, we display the residual azimuthal variations in oxygen, europium, iron, and barium abundance at the present day, calculated at distances of 4, 6, and 9 kpc.

It is worth underlining the presence of   additional wiggles in the azimuthal variations compared to the results of Model A (see Fig. \ref{residual_A}, where the single mode  $m=2$ was imposed), which arise from the coexistence and interplay of different modes. However, the amplitude of the azimuthal variation at 9 kpc remains roughly the same. Therefore, in the subsequent section, we will employ Model A as our reference model for further investigations.

\begin{figure*}
\centering
\includegraphics[scale=0.3]{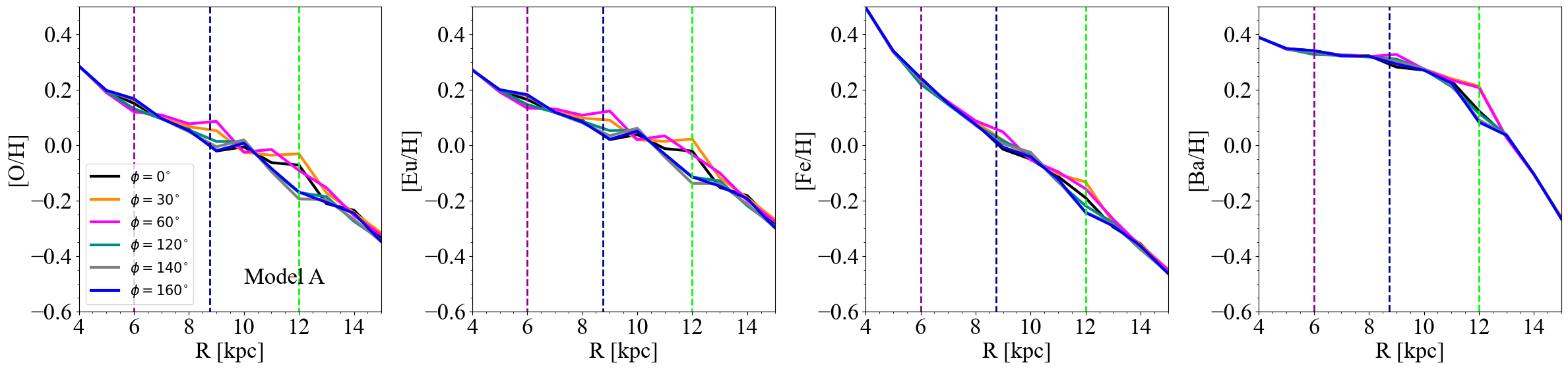}
 \includegraphics[scale=0.3]{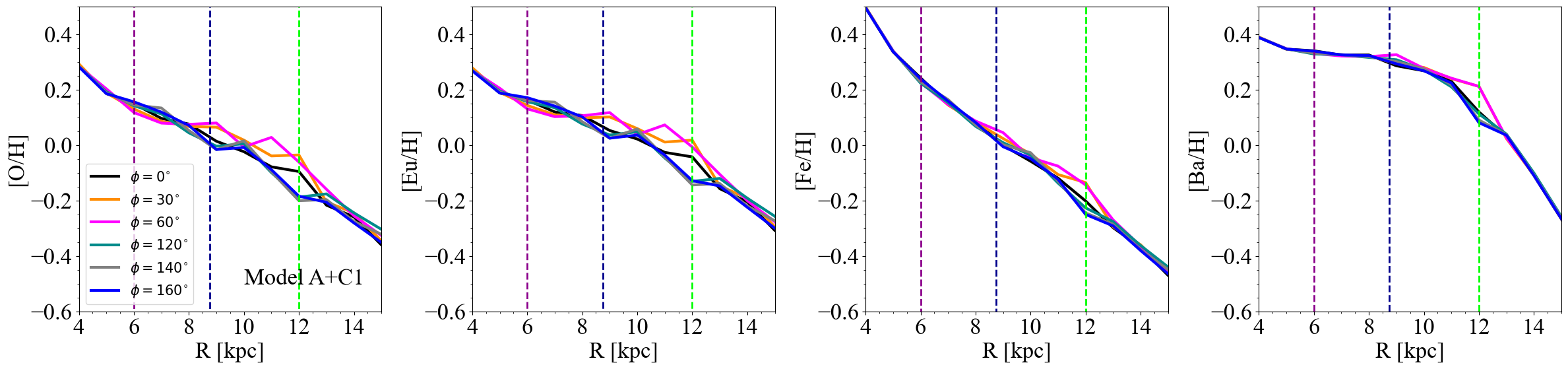}
  \includegraphics[scale=0.3]{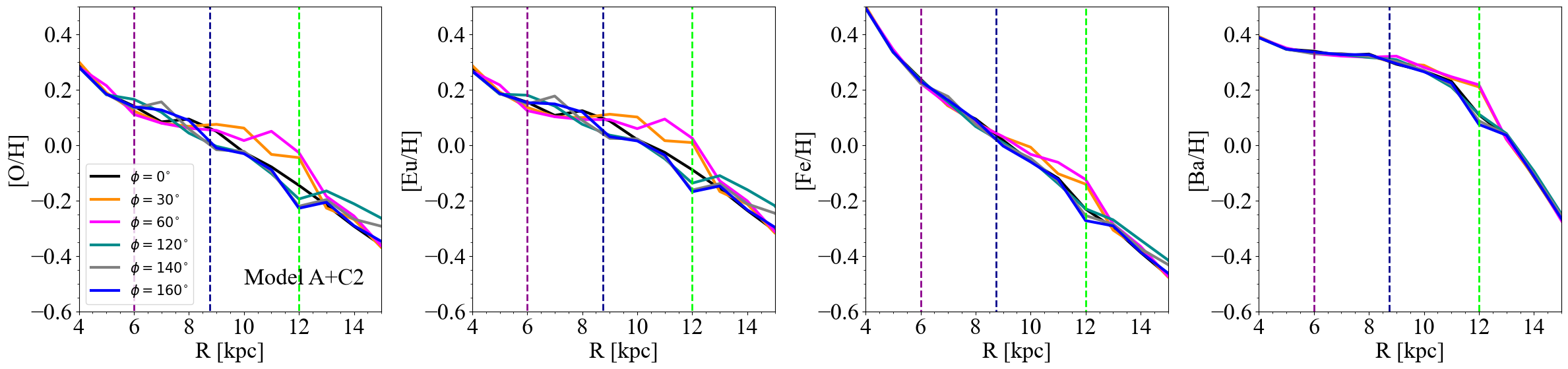}
   \includegraphics[scale=0.3]{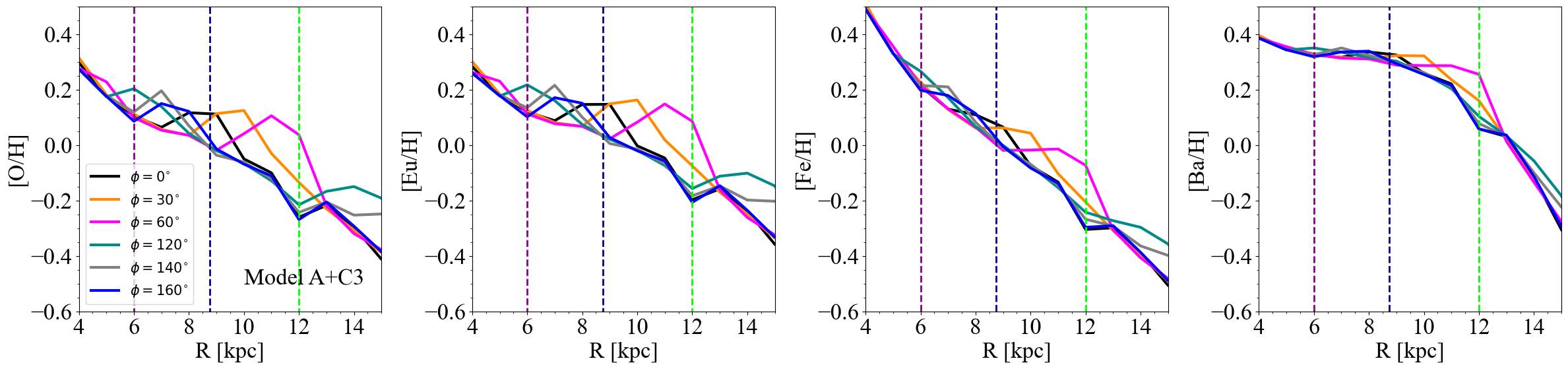}
  \caption{Comparing azimuthal variations in abundance gradients  for oxygen, europium, iron and barium  predicted by Model A (first row) with models where  the condition of transient spiral have been considered.
  Model A+C1 where in the last 100 Myr of Galactic evolution the condition $\Omega_{s,j}(R)=\Omega_d(R)$ is valid for all the Galactic radii (co-rotation extended at all distances) is reported in the second row, Model A+C2 when the above-mentioned condition lasted for the last 300 Myr in the third row, Model A+C3 for 1 Gyr in the last row. Colour convention of the lines  as in Fig. \ref{gradient_A}. The vertical lines indicate the  corotations of Model A.  }
\label{fig_C1}
\end{figure*}

\begin{figure*}
\includegraphics[scale=0.25]{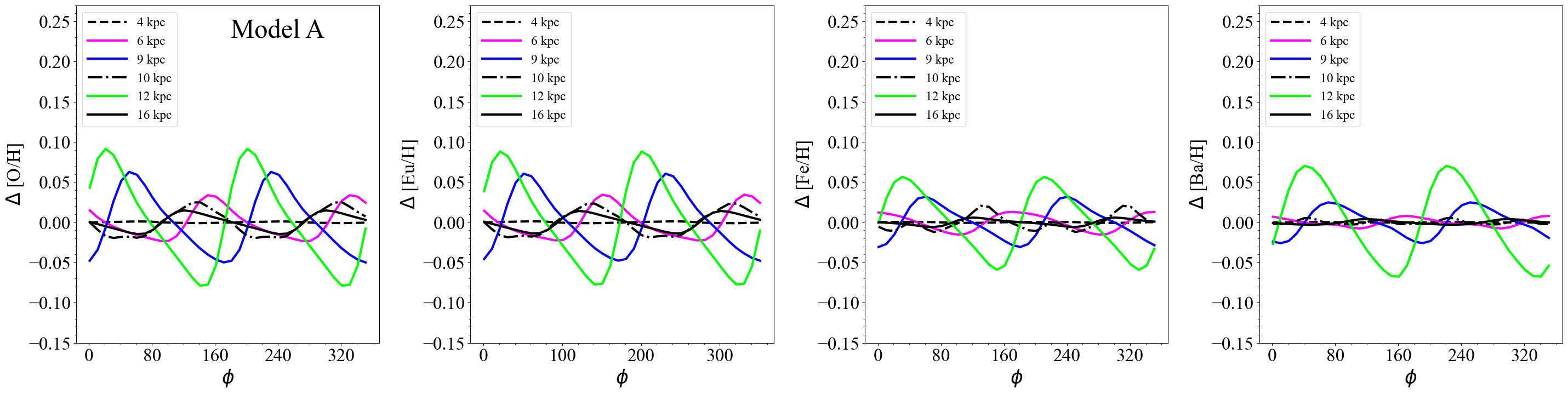}
\includegraphics[scale=0.25]{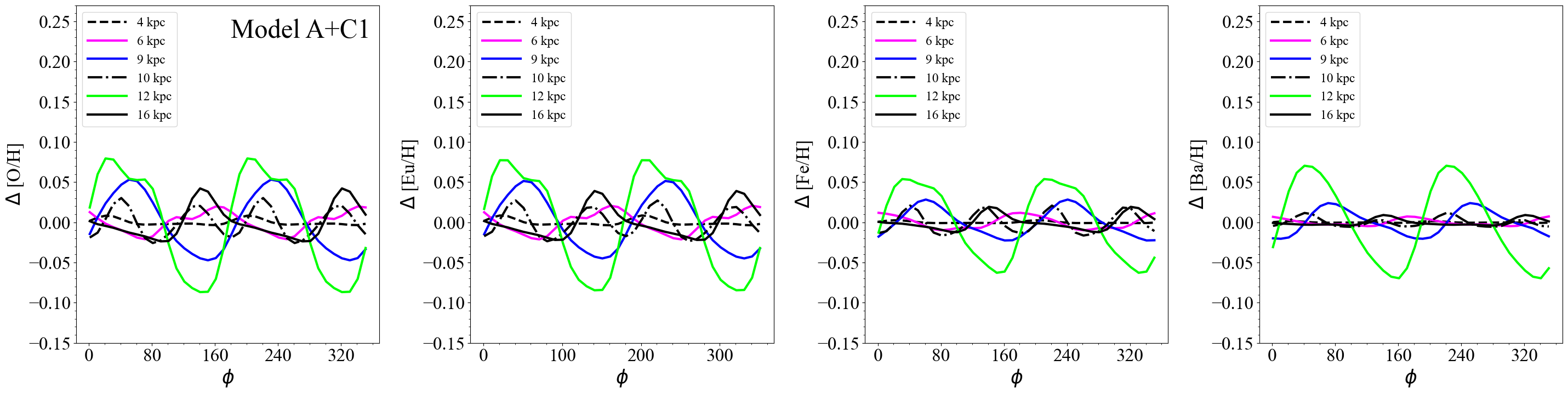}
\includegraphics[scale=0.25]{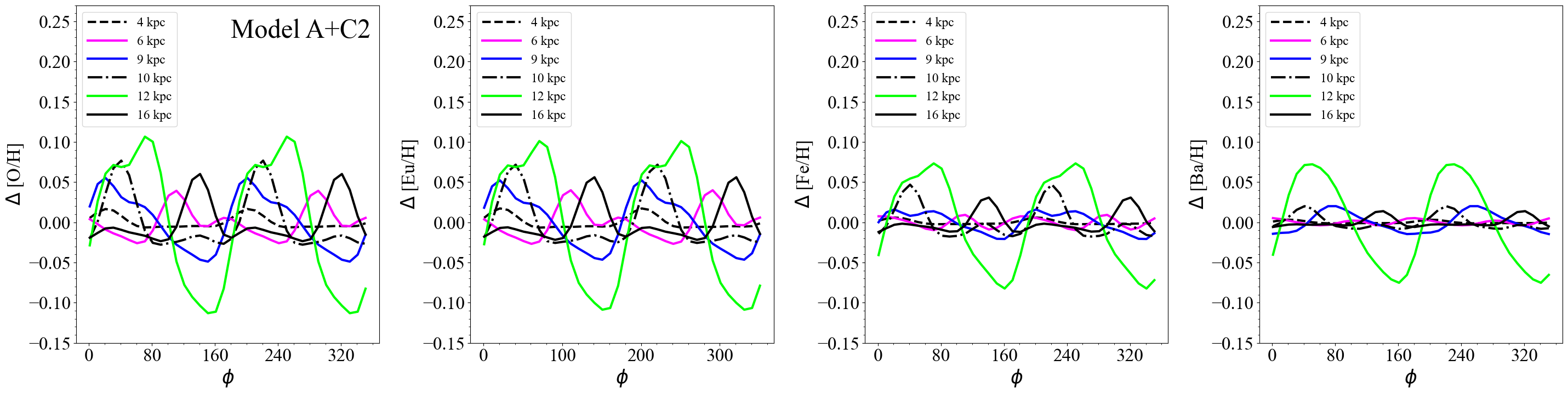}
\includegraphics[scale=0.25]{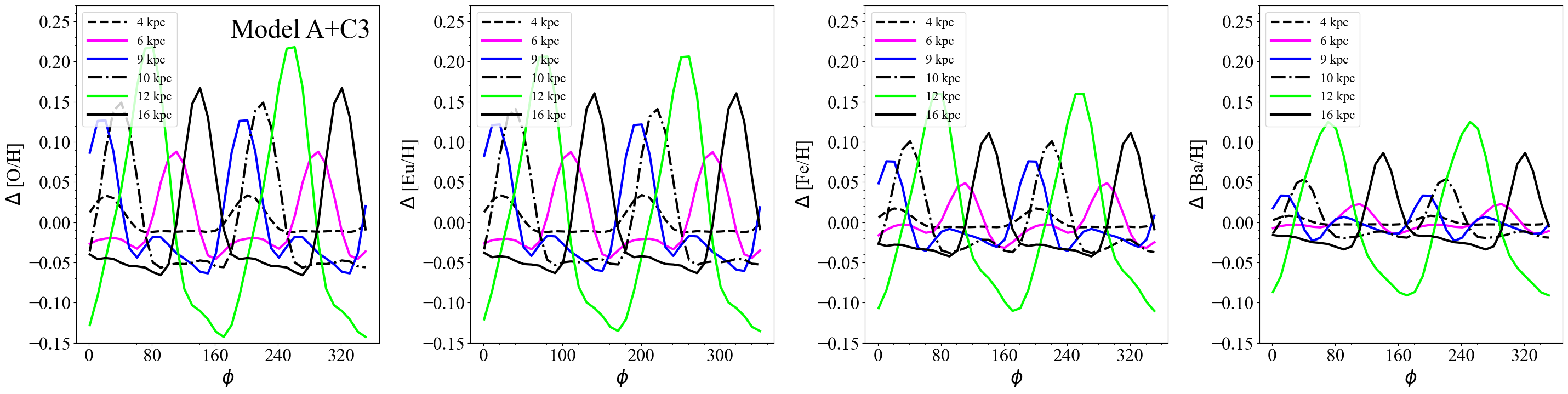}
\centering
   \caption{ As in Fig. \ref{residual_A}, but also for Model A+C1 (second row), Model A+C2 (third row) and Model A+C3 (last row), respectively.
 }
\label{D_AC1}
\end{figure*}

\begin{figure*}
\centering
\includegraphics[scale=0.3]{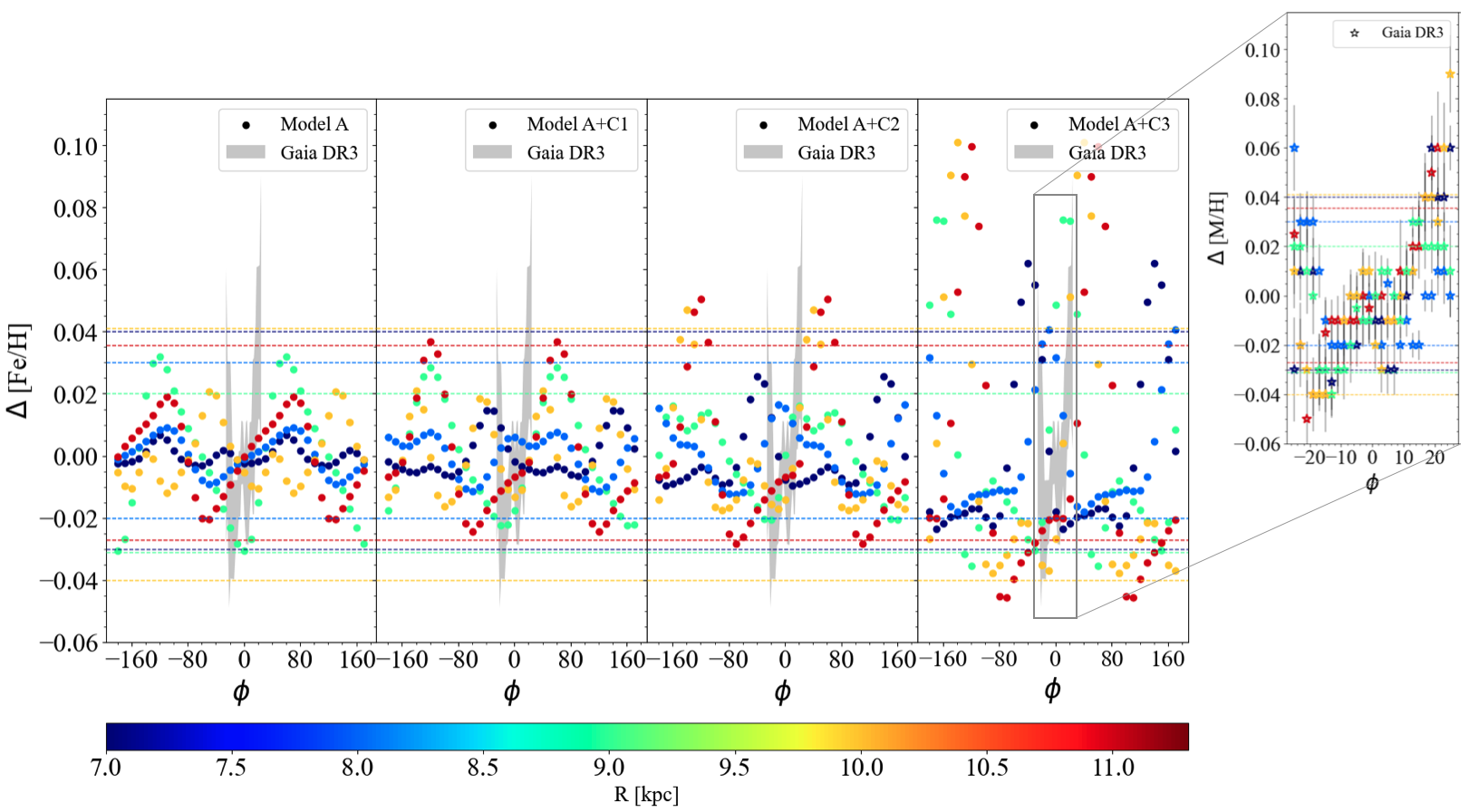}
 \caption{ Present-day residual azimuthal variations in  iron respectively computed at different Galactocentric distances for the multiple spiral structure of the Model A, A+C1, A+C1 and A+C3, compared with metallicity variation found by \citet{poggio2022} analysing  GSP-Spec abundances of \gaia DR3 \citet{recioDR32022b} (grey shaded regions). 
 Horizontal lines with the same colour indicate the 10$\%$ and 90 $\%$ percentiles of the metallicity variation as computed by \citet{poggio2022} at different Galactocentric distances.  In each panel, the shaded grey area indicates the region spanned by \citet{poggio2022} data.  In the zoom-in plot associated with the fourth panel, we report the \citet{poggio2022} data   indicating median metallicity for Sample A as a function of Galactic azimuth  for different Galactocentric distances (after the median metallicity of the stars for each ring has been subtracted).  }
\label{poggio}
\end{figure*}

\begin{figure}
\centering
\includegraphics[scale=0.37]{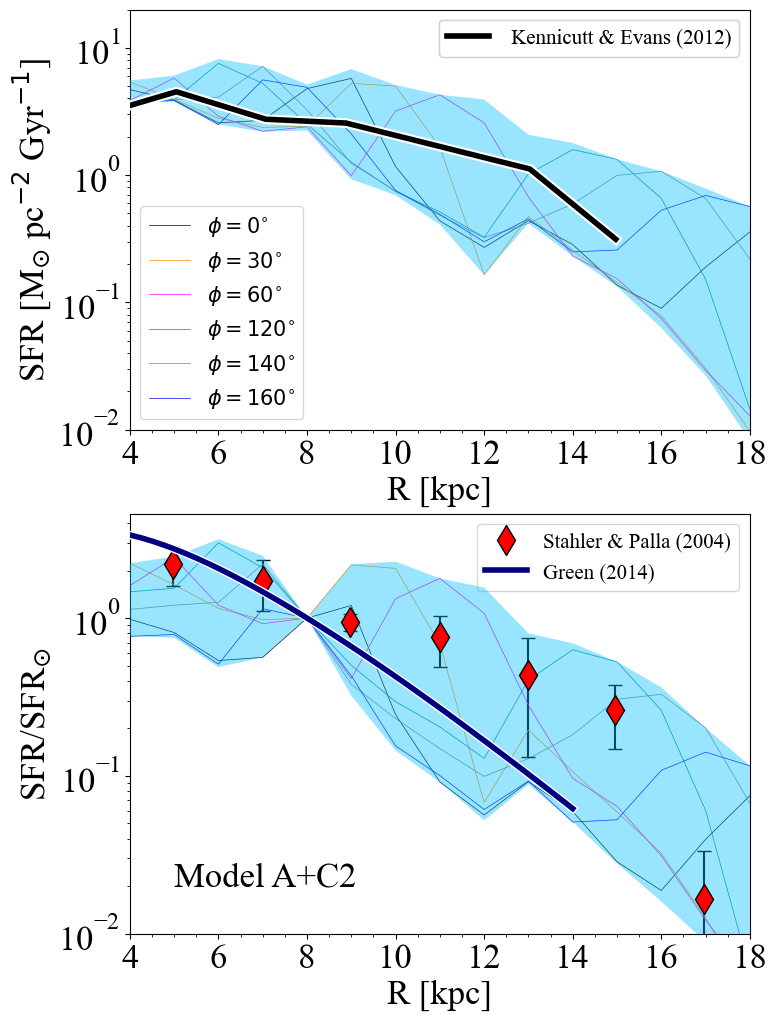}
 \caption{As in Fig. \ref{SFR_gradientA} but for the Model A+C2. }
\label{gradient_SFR_AC2}
\end{figure}

\begin{figure*}
\centering
\includegraphics[scale=0.25]{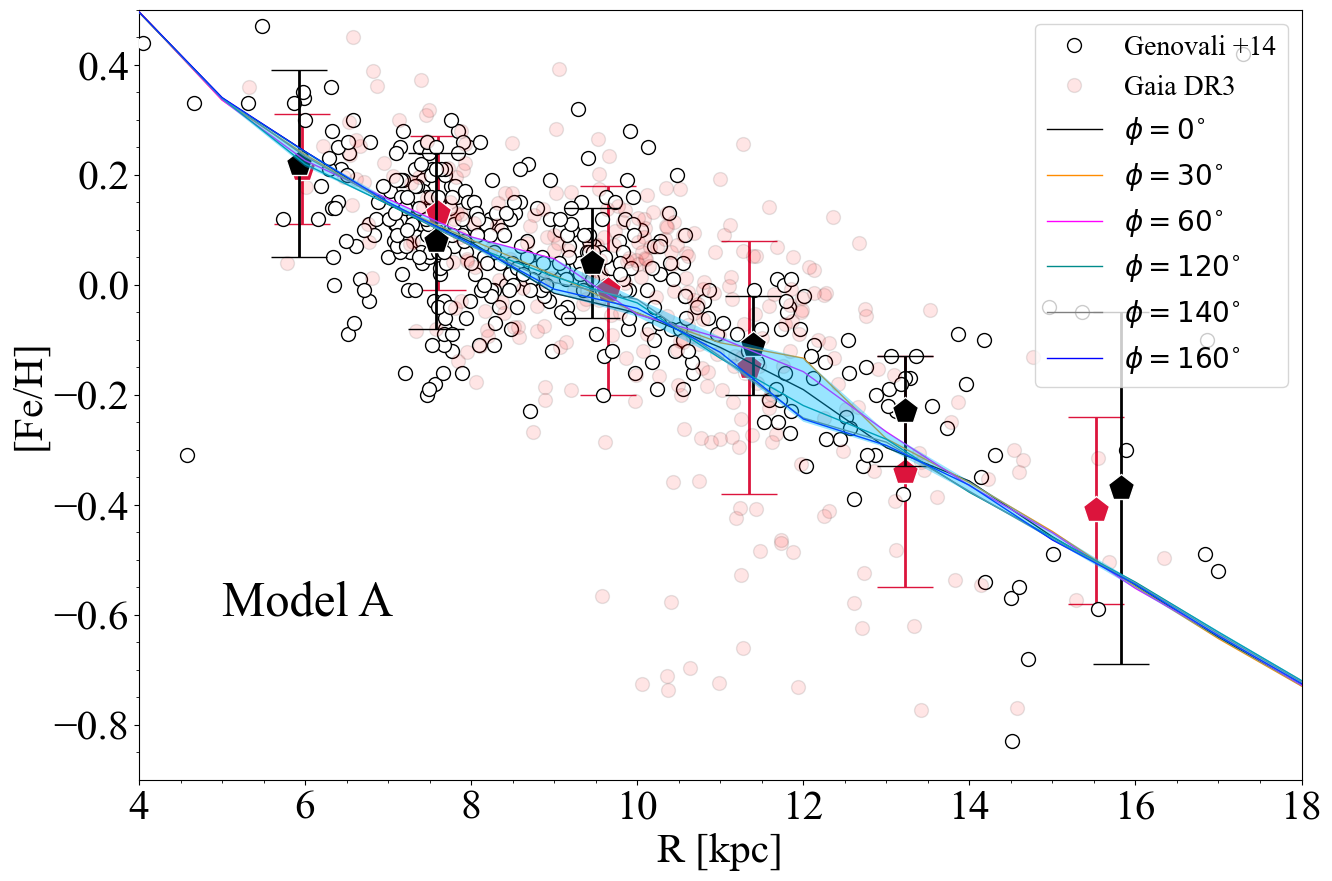}
\includegraphics[scale=0.25]{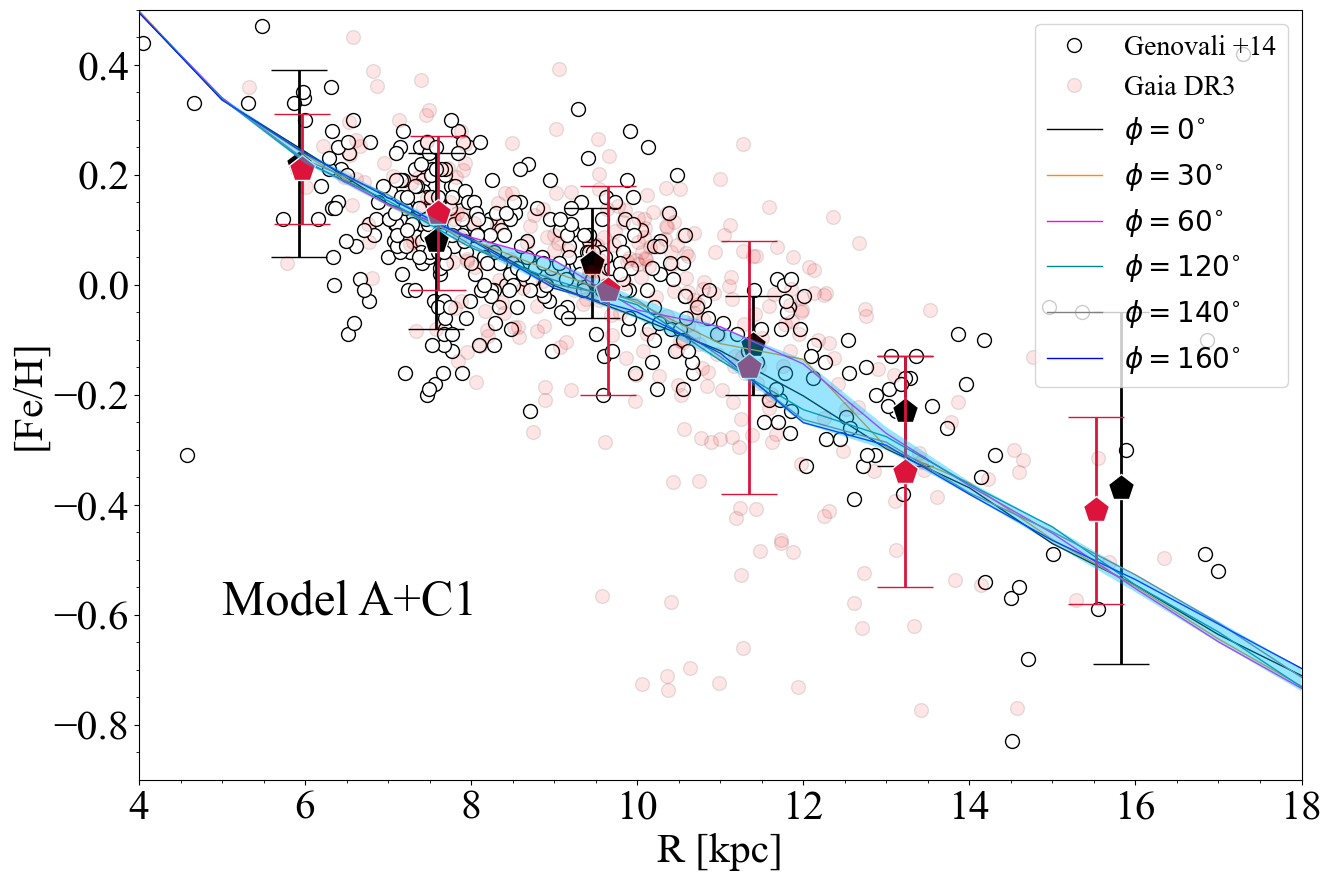}
\includegraphics[scale=0.25]{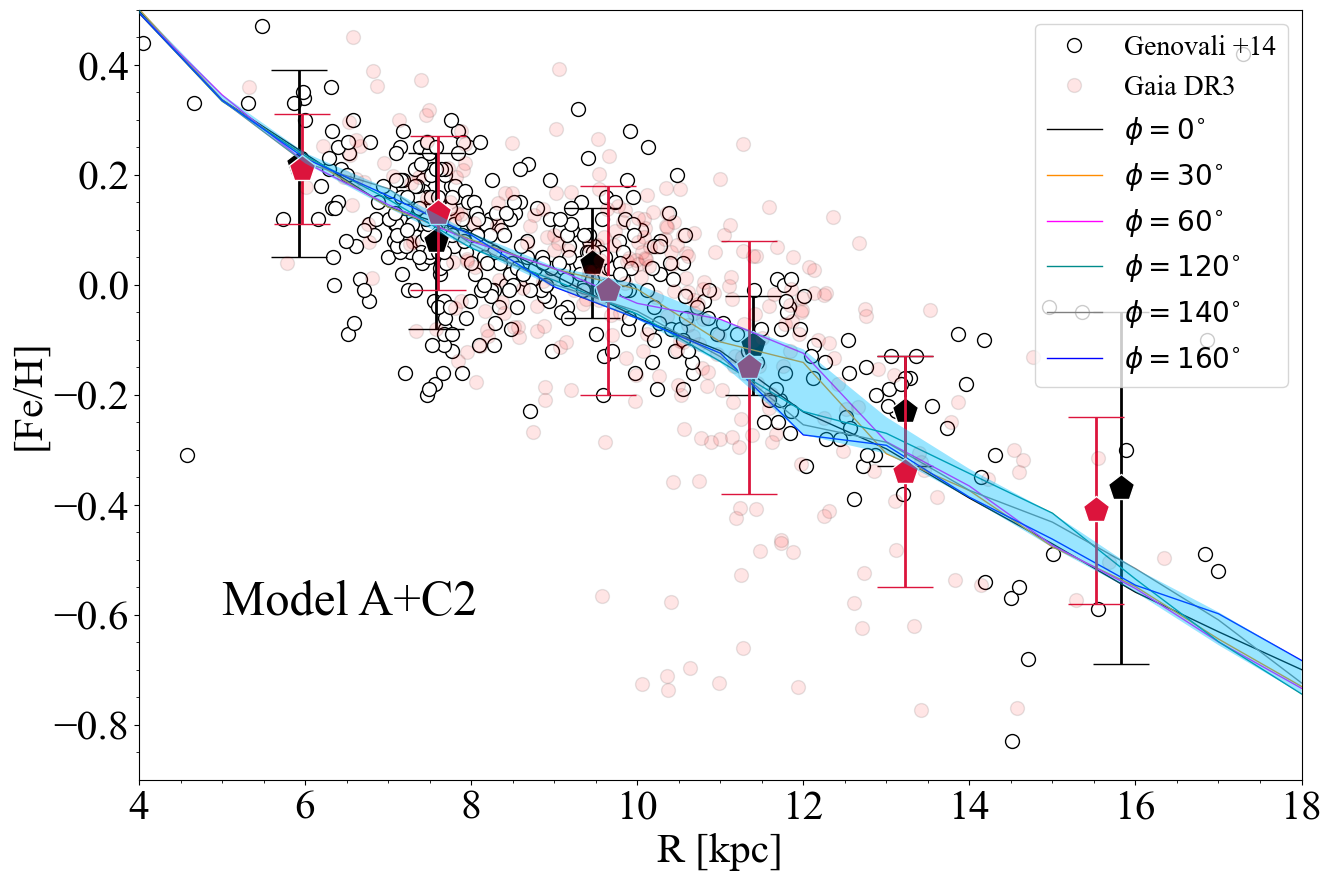}
\includegraphics[scale=0.25]{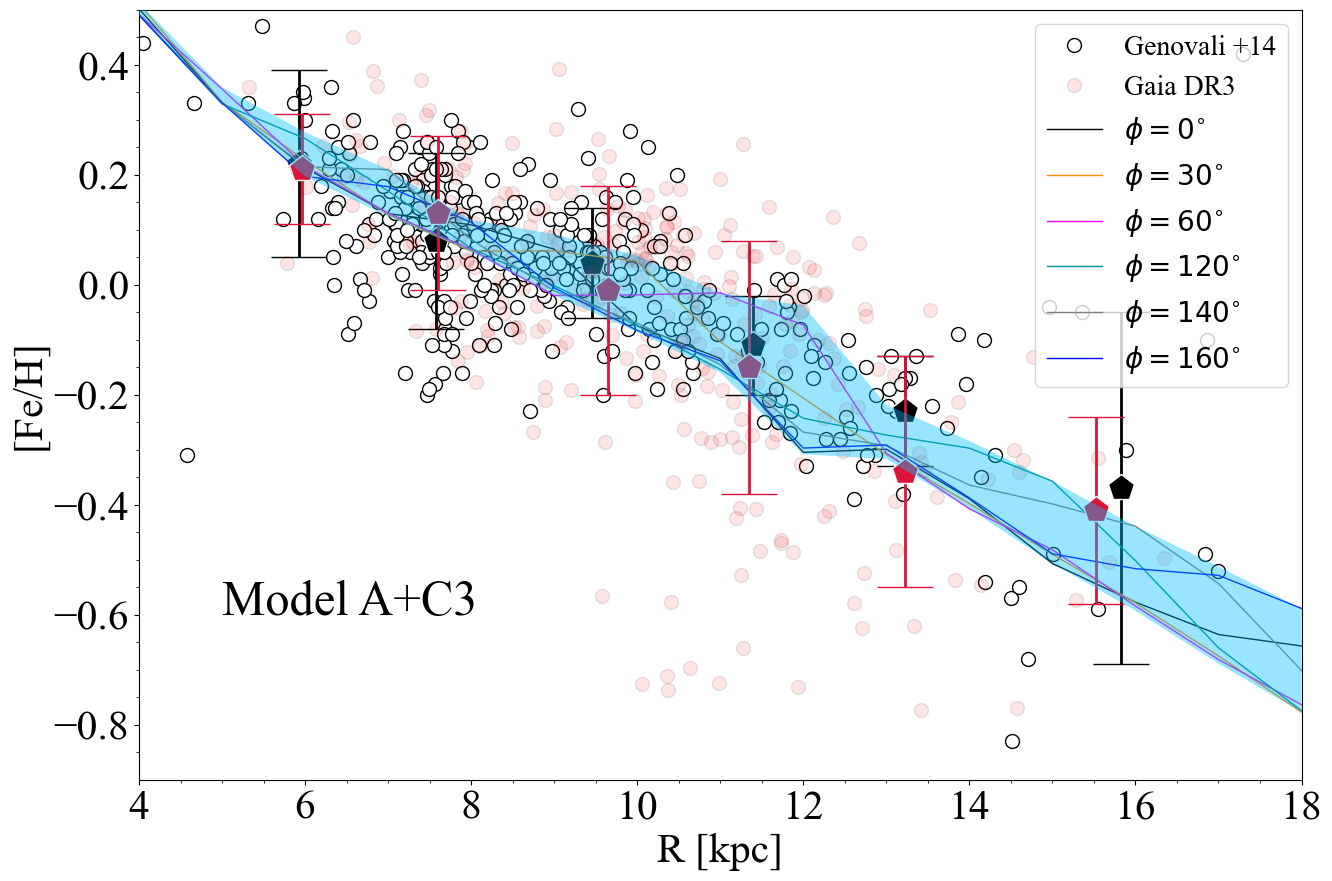}
 \caption{Present-day iron  abundance gradient predicted  by Models A, A+C1, A+C2 and A+C3 at different azimuthal coordinates.  In each panel, the shaded light-blue area denotes the range of maximum and minimum Fe abundance values at various distances from the centre of the galaxy spanned by the models. The Cepheids  data collected by \citet{genovali2014} and  GSP-Spec [M/H] for \gaia DR3 source id of  Cepheids of \citet{ripepi2022} are reported with empty   and full light-red dots, respectively. The average abundance values and relative errors, when dividing these two data sets into six radial bins, are shown by black and red pentagons, respectively.    }
\label{cepheids}
\end{figure*}

 \subsection{Extending the  co-rotation to all Galactocentric distances (Models A+C1, A+C2,  A+C3) }
\label{results_C}

The results of chemical evolution models presented so far are based on the evidence that the Milky Way  can contain multiple patterns, with slower patterns situated towards outer radii \citep{minchev2006,quillen2011}.  Furthermore,  in other dynamical works \citep{grand2012,hunt2019}   co-rotating arms can be found at all radii  in transient spiral arm structures.

The occurrence of temporary spiral structure results in phase mixing, leading to the formation of ridges and arches in the nearby kinematics observed in the \gaia data \citep{recioDR32022b,antoja2021,ramos2018,palicio2023}. By incorporating a bar with various pattern speeds, \citet{grand2012,hunt2019}   successfully generated a reasonably accurate resemblance to the observations. 
In addition,  in the pure N-body  simulation of \citet{grand2012} of a barred galaxy,   the spiral arms are transient features  whose pattern speeds decline with the radius, so that the pattern speed closely matches the rotation of star particles. 

To mimic the above-mentioned scenarios, we impose that at recent  evolutionary times, the pattern speeds $\Omega_{s,j}$ match the rotational curve at all radii, i.e the condition $\Omega_{s,j}(R)=\Omega_d(R)$ is  extended to all Galactocentric distances for each of the spiral chunks  of Model A. 
As indicated in Table \ref{tab_models},  we applied this condition for the last 100 Myr of Galactic chemical evolution  (Model A+C1),  300 Myr of evolution (Model A+C2) and  1 Gyr (Model A+C3), respectively.

In Figs. \ref{fig_C1} and \ref{D_AC1}, we presented the abundance gradients and the residual azimuthal variations predicted by these three new models.
As expected, the amplitudes  for the considered chemical elements are amplified compared with Model A, especially at  Galactocentric distances different from that of  the corotations of Model A, as visible in Figs. \ref{fig_C1} for  Models A+C1 and A+C2. The longer the condition for the transient spiral arm, the larger the amplitude of the azimuthal variation. In fact, as discussed in previous Sections, the condition $\Omega_{s,j}(R)=\Omega_d(R)$ imposes that,  at  fixed  Galactocentric distance $R$ and azimuthal coordinate $\phi$, the perturbation term $\delta_{MS}$ (introduced in eq.  \ref{delta2})  in the SFR  does not vary in time. Hence, the chemical fluctuation  should be  amplified. 
The extreme case of Model A+C3 with the condition $\Omega_{s,j}(R)=\Omega_d(R)$    in the last 1 Gyr, should be considered as a test (see last rows in Figs. \ref{fig_C1} and \ref{D_AC1}).   For instance, for oxygen, the maximum amplitude at 12 kpc  is $\Delta_{max}$[O/H]$\sim$ 0.37 dex,  whereas for Model A was $\Delta_{max}$[O/H]$\sim$ 0.17 dex, hence increased by a factor of 2.18.

\subsubsection{Tightly wound spiral structures}

The  work of \citet{quillen2018} and \citet{laporte2019} suggested that tightly wound spiral structures should be considered based on the modelling of
phase-space structure found in
the second Gaia data release   \citep{katz2018}.
A smaller pitch angle gives rise
to a more  tightly wound spiral structure. In \citet{reshe2023}, they study  pitch angles of spiral arms in galaxies within the Hubble Space Telescope COSMOS field.  Analysing a  sample of 102 face-on galaxies with a two-armed pattern they found a  decreasing trend in the pitch angle value from a redshift range of $z=1$ to $z=0$. 
However, in this study, we do not test the effects of a decreasing pitch angle in time on the chemical evolution of the Galactic disc. In fact, as already pointed out by ES19, the amplitude of the  azimuthal variation in the abundance gradients is not dependent on the pitch angle. As highlighted by Fig. 18 of ES19,  small pitch angles solely   reduce the phase difference  of the abundance variation between different radii.

\section{Gaia DR3 data}
\label{gaia_sec}
  \gaia DR3 \citep{vallenari2022} and \citet{recioDR32022a,recioDR32022b} have brought a truly and unprecedented revolution opening a new era of all-sky spectroscopy. 
With about 5.6 million stars, the \gaia DR3 General Stellar Parametrizer - spectroscopy   (GSP-Spec, \citealt{recioDR32022a}) all-sky catalogue is the largest compilation of stellar chemo-physical parameters and the first one from space data without the issues of biased samples which hampered the observations from Earth. 
In  \citet{recioDR32022b}, the high quality of the GSP-Spec chemical abundances for $\alpha$-elements and Fe have been used  to provide important constraints on the Galactic  Archaeology.  Updated chemical evolution models for the evolution of the thick and thin discs have been presented by \citet{spitoni2023} constrained by $\alpha$-elements in the solar vicinity. The interaction with the Sagittarius dwarf galaxy could explain the  observed feature in the abundance ratios. In  \citet{contursi2023}, they analysed the GSP-Spec  cerium   unveiling the  evolution history of this heavy element.

In Section \ref{poggio_sec} we compare our model predictions with the azimuthal variations found in the metallicity  distribution of \citet{poggio2022}, whereas in Section \ref{sec_ceph} we compare them  with the  GSP-Spec [M/H] abundance ratios of Cepheids.

\subsection{Comparison with \citet{poggio2022}}\label{poggio_sec}

\citet{poggio2022} exploited  \gaia DR3 data providing a map of inhomogeneities in the Milky Way's disc [M/H] abundances, which extends to approximately 4 kpc from the solar position. This was achieved by studying various samples of bright giant stars, which were selected based on their effective temperatures and surface gravities using the GSP-Spec module. Their Sample A,  composed  of hotter  (and younger) stars, exhibits significant inhomogeneities, which manifest as three (possibly four) metal-rich elongated features that correspond to the spiral arms' positions in the Galactic disc.  In Fig. \ref{poggio},  we compare   the  present-day residual azimuthal variation $\Delta$[Fe/H] predicted by Models A, A+C1, A+C2 and A+C3 with $\Delta$[M/H] of \citet{poggio2022}   Sample A because they   should  better trace the present-day ISM inhomogeneities predicted by our models. 
We recall that in GSP-Spec [M/H] values follow the [Fe/H] abundance with a tight correlation. For this reason,  in the following plots, we compare [Fe/H] ratios predicted by our models with  Gaia DR3 [M/H] abundance ratios.

We note that the amplitude of the variations predicted by Model A is smaller than the one  displayed by the Sample A  of \citet{poggio2022}  in  the interval between the
10$\%$ and 90$\%$ percentiles. It is also important to stress that in \citet{poggio2022} data, there is not a strong dependence of the amplitude with the radius in contrast with Model A results. In fact, in the range of Galactocentric distances of  Fig. \ref{poggio}, Model A shows the maximum  amplitude of the azimuthal fluctuation near the  co-rotation radius of the second spiral chunk ($\Omega_{s,2}$ =20 km s$^{-1}$ kpc$^{-1}$)  and almost negligible azimuthal variation is found at 7 kpc.

However, the agreement is quite good with  Models A+C1, A+C2 where for the last 100 Myr and 300 Myr, respectively of evolution we impose the transient spiral arm condition (i.e. $\Omega_{s,j}(R)=\Omega_D(R)$). On the other hand, Model A+C3 produces azimuthal variations much larger than the observed ones. 
In Fig. \ref{gradient_SFR_AC2}, we show that the present-day SFR profile throughout the Galactic disc for Model A+C2 is in  agreement with observations. Compared to Fig. \ref{SFR_gradientA}, in this case we see that slightly higher  peaks of SF are predicted in the Galactic region enclosed between 9 and 12 kpc.

\subsection{Comparison with \gaia DR3 GSP-Spec Cepheids }\label{sec_ceph}

\citet{ripepi2022}  presented the Gaia DR3 catalogue of Cepheids of all types, obtained through the analysis carried out with the Specific Object Study (SOS) Cep$\&$RRL pipeline.
In Fig. \ref{cepheids}, we show the abundance gradient of  \gaia DR3 Cepheids in the Galactic disc with calibrated GSP-Spec metallicity [M/H]  as  suggested by  \citet{recioDR32022a}. 
We impose that the fluxNoise  flag of Table 2  \citet{recioDR32022a}  has been set equal to 0 (best quality data). The \gaia source id of the Cepheid sources  are those identified in \gaia DR3 \citep{ripepi2022}.
We computed the Galactocentric distances by adopting the Sun’s Galactocentric position $(R,Z)_{\odot}=(8.249,0.0208)$ kpc \citep{gravity2021,bennett2019} 
and the high-precision astrometric parameters from \gaia EDR3
\citep{brown2021} and the  additional  information provided by  \gaia DR3 for the radial velocities \citep{katz2022,vallenari2022}. 

In Fig. \ref{cepheids}, we note that the abundance gradient emerged by GSP-SPec metallicity is in good agreement with the ones of \citet{genovali2014}.  The larger spread in the GSP-SPec metallicity is due to the higher mean uncertainties ($\sim 0.15$ dex) compared to the ones computed with high-resolution spectroscopy in \citet[][$\sim 0.08$ dex]{genovali2014}.  
In the same plots, we also highlighted the abundance variation in [Fe/H] predicted by  Models A, A+C1, A+C2, and A+C3. We note that the only model which partially can account for the spread in \gaia data and in \citet{genovali2014} is Model A+C3. We recall that this last case should be considered as an extreme case where the all-corotation radii condition lasted for 1 Gyr.  In conclusion, we believe that the observed spread in the abundance gradient is only partially explainable through spiral arms and other dynamical processes should be taken into account.

\section{Conclusions and Future perspectives}
\label{conclu_sec}
In this paper, we presented an updated version of the 2D  chemical evolution model for the Galactic disc presented by \citet{spitoni2D2018} considering the density fluctuation created by multiple pattern spiral arms. We studied in detail their effects on the abundance gradients of oxygen, iron, barium and europium. In particular, for the predicted [Fe/H] we also show the comparison with the recent GSP-Spec [M/H] abundances \citep{recioDR32022b, poggio2022}.
The main results can be summarised as follows:
\begin{itemize}

\item That azimuthal
variations  are dependent on the considered chemical element. Elements synthesised   on short time scales (i.e., oxygen and europium
in this study) exhibit larger abundance  fluctuations. In fact, having progenitors with short lifetimes, the chemical elements restored in the ISM trace perfectly  the star formation rate  perturbed by the passage of  spiral arms. It results in important azimuthal variations of the  abundance gradient compared with other elements ejected into the ISM with a significant delay (i.e., iron and barium).

\item The  2D map of the projected star formation rate onto the Galactic disc in the presence of spiral arms the multiple patterns predicted by the Model A  (see Table \ref{tab_models}) presents arcs and arms compatible with tracers of spiral arms (young UMR stars, Cepheids, distribution of stars with low radial actions).
\item  
 As found by \citet{spitoni2D2018} in the study of single pattern spiral arms,  the largest fluctuations in the azimuthal abundance gradients are found near the co-rotation radius where the relative velocity, with respect to the disc, is close to zero.  Larger azimuthal variations are associated with the most external spiral clumps where the associate   co-rotation radius is placed at  larger Galactocentric distances.
\item 
Assuming  that the modes with different patterns combine linearly, we showed that also the total effects of the different modes  on abundance azimuthal variations respond linearly to different  modes considered.

\item Imposing that  the pattern speeds match the Galactic rotational curve at all radii,  in the last 100 Myr of evolution, has the effect to amplify the azimuthal variation.

\item  Predicted azimuthal variation are consistent with  metallicity variations found by  \citet{poggio2022}  \gaia DR3, if transient spiral arms are assumed at recent evolution times (during the last $\simeq 300$ Myr).

\end{itemize}

In the future, we plan to explore the scenario where the spiral pattern winds up at $\Omega_d = \Omega(R) - \kappa/2$ (where $\kappa$ is the epicyclic frequency), as proposed by \citet{bland2021}.
\citet{hunt2019} highlighted the intricate challenge of separating the impacts of the bar and the spiral structure. Hence, in future work, we plan to include also variations produced by the Galactic bar \citep[e.g.][]{palicio2018,palicio2020}.
In Barbillon et al. (in prep.), we would like to extend the analysis of \citet{poggio2022} to other  GSP-Spec chemical elements (i.e. total $\alpha$, Mg, Ca, Si, Ti) and compare them with our models.
 We plan also to consider stellar migration as an additional dynamical process in our model. In fact, several works in a cosmological context highlighted the importance of stellar migration in the azimuthal variation of abundance gradients in the vicinity of spiral arms \citep{grand2012,grand2014,grand2016,sanchez_Me2016}.
 It is also our intention to study  the effects of spiral arms on the Galactic chemical evolution of short-lived radionuclides, such as $^{26}$Al and $^{50}$Fe using the same  model and nucleosynthesis prescriptions as in \citet{vasini2022,vasini2023}. Because  these elements are  tracers of the star formation, we expect the signature of the passage of  spiral structures  on their present-day distribution \citep{siegert2023}. 

 In the future, we plan to test in our chemical evolution model  the effects   of  gas flows at the co-rotation as highlighted in \citet{barros2021}.

\section*{Acknowledgement}

 The authors thank the anonymous referee for various suggestions that improved the paper.
We  thank P. De Laverny,  S. Khoperskov and M. Sormani for  useful discussions.
E. Spitoni and A. Recio-Blanco received funding from the European Union’s Horizon 2020 research and innovation program under SPACE-H2020 grant agreement number 101004214 (EXPLORE project). 
This project has received funding from the European Union's Horizon 2020 research and innovation programme under the Marie Sklodowska-Curie grant agreement N. 101063193.
This work has made use of data from the European Space Agency (ESA) mission
\gaia (\url{https://www.cosmos.esa.int/gaia}), processed by the \gaia
Data Processing and Analysis Consortium (DPAC,
\url{https://www.cosmos.esa.int/web/gaia/dpac/consortium}). Funding for the DPAC
has been provided by national institutions, in particular the institutions
participating in the \gaia Multilateral Agreement.
 I. Minchev acknowledges support by the Deutsche Forschungsgemeinschaft under the grant MI 2009/2-1. 
P. A. Palicio acknowledges the financial support from the Centre national d’études spatiales (CNES).
This work was partially supported by the European Union (ChETEC-INFRA, project no. 101008324). F. Matteucci and A. Vasini  thank I.N.A.F. for the 1.05.12.06.05 Theory Grant
- Galactic archaeology with radioactive and stable nuclei.

\bibliographystyle{aa} 
\bibliography{disk}

\end{document}